\documentclass[aps,prd,twocolumn,superscriptaddress,nofootinbib]{revtex4-1}

\usepackage{latexsym}
\usepackage{amsmath}
\usepackage{amssymb}
\usepackage{amsfonts}
\usepackage{bm}

\usepackage{color}

\usepackage{supertabular} 
\usepackage{placeins}
\usepackage{epsfig}
\usepackage{graphicx}

\begin{document}

%\preprint{}

\title{$QQ\bar{s}\bar{s}$ tetraquarks in the chiral quark model}


\author{Gang Yang }
\email[]{yanggang@zjnu.edu.cn}
%\thanks{}
\affiliation{Department of Physics, Zhejiang Normal University, Jinhua 321004, China}

\author{Jialun Ping}
\email[]{jlping@njnu.edu.cn}
\affiliation{Department of Physics and Jiangsu Key Laboratory for Numerical Simulation of Large Scale Complex Systems, Nanjing Normal University, Nanjing 210023, P. R. China}

\author{Jorge Segovia}
\email[]{jsegovia@upo.es}
\affiliation{Departamento de Sistemas F\'isicos, Qu\'imicos y Naturales, \\ Universidad Pablo de Olavide, E-41013 Sevilla, Spain}


%%%%%%%%%\date{\today}

\begin{abstract}
The low-lying $S$-wave $QQ\bar{s}\bar{s}$ ($Q=c, b$) tetraquark states with $IJ^P=00^+$, $01^+$ and $02^+$ are systematically investigated in the framework of complex scaling range of chiral quark model. Every structure including meson-meson, diquark-antidiquark and K-type configurations, and all possible color channels in four-body sector are considered by means of a commonly extended variational approach, Gaussian expansion method. Several narrow and wide resonance states are obtained for $cc\bar{s}\bar{s}$ and $bb\bar{s}\bar{s}$ tetraquarks with $IJ^P=00^+$ and $02^+$. Meanwhile, narrow resonances for $cb\bar{s}\bar{s}$ tetraquarks are also found in $IJ^P=00^+$, $01^+$ and $02^+$ states. These results confirm the possibility of finding hadronic molecules with masses $\sim\,0.6\,\text{GeV}$ above the noninteracting hadron-hadron thresholds.
\end{abstract}


\maketitle

%%%%%%%%%%%%%%%%%%%%%%%%%%%%%%%%%%%%%%%%%%%%%%%%%%%%%%%%%%%%%%%%%%%%%%%%%%%%%%%%

\section{Introduction}

We are witnessing in the last two decades of a big experimental effort for understanding the heavy-flavor quark sectors of both meson and baryon systems. Many experiments have been settled worldwide such as B-factories (BaBar, Belle, and CLEO), $\tau$-charm facilities (CLEO-c and BES) and hadron-hadron colliders (CDF, D0, LHCb, ATLAS, and CMS) providing a sustained progress in the field with new measurements of conventional and exotic heavy-flavored hadrons. 

Within the baryon sector, and attending mostly to the spectrum, five excited $\Omega_c$ baryons were announced three years ago by the LHCb collaboration in the $\Xi_c^+ K^-$ mass spectrum~\cite{lhcb:2017omegac} and, very recently, the same collaboration has reported additional four narrow excited states of the $\Omega_b$ system in the $\Xi_b^0 K^-$ mass spectrum~\cite{lhcb:2020omegab}. In 2019, two excited bottom baryons, $\Lambda_b^0(6146)$ and $\Lambda^0_b(6152)$, were discovered in the LHCb experiment~\cite{lhcb:2019lambdab1}. Later on, the LHCb collaboration also announced one more $\Lambda^0_b$ baryon around 6070 MeV in the $\Lambda^0_b\pi^+\pi^-$ invariant mass spectrum~\cite{lhcb:2019lambdab2}, which is consistent with the reported one of the CMS collaboration~\cite{cms:2020lambdab}. Additionally, three excited $\Xi_c^0$ states were announced by the LHCb collaboration in the $\Lambda^+_c K^-$ mass spectrum~\cite{lhcb:2020Xic}. 

All of these newly discovered baryons undoubtedly complement the scarce data on heavy flavor baryons in the Review of Particle Physics (RPP) of the Particle Data Group~\cite{Tanabashi:2018oca}. Furthermore, these experimental findings trigger a large number of theoretical investigations. The three-quark structure of the new $\Omega_c$ baryons has been claimed by QCD sum rules~\cite{hxc:2017dpp} and different potential models~\cite{mkjlr:2017vneb, klw:2017unosttd, gy:2018dsichqm}. Also, the description of the $\Omega_b$ signals as $P$-wave conventional baryons is preferred by phenomenological quark model approach~\cite{mkjl:2020ies, lyx:2020pinob}, heavy quark effective theory~\cite{hxc:2020ebfssi} and QCD sum rules~\cite{zgw:2020atwqsr}. Meanwhile, the baryon-meson molecular interpretation has been suggested for the excited $\Omega_b$ baryons in Ref.~\cite{whl:2020osmbms}. The $\Lambda^0_b(6072)$, $\Lambda^0_b(6146)$ and $\Lambda^0_b(6152)$ have been identified as radial and angular excitations within QCD sum rules~\cite{ka:2020nsa2sbb, klw:2019inolsicqm, bc:2019iolsa1dbb, hmy:2020ixboblapwxb} and chiral quark models~\cite{klw:2020unosttd, qfl:2020citnoxr}. However, the $\bar{D}\Lambda-\bar{D}\Sigma$ molecular configurations have been also suggested for these states in Ref.~\cite{hqz:2020dtnoxsams}.

Apart from \emph{conventional} heavy flavored baryons, there are limited results on open-bottom mesons and detailed studies of the open-charm ones were not undertaken until large datasets were obtained by CLEO at discrete energy points and by the B-factory experiments using radiative returns to obtain a continuous exposure of the mass region. The picture that has emerged is complex due to the many thresholds in the region. This resembles the experimental situation found in the heavy quarkonium spectrum with the observation of more than two dozens of unconventional charmonium- and bottomonium-like states, the so-called XYZ mesons. However, still successful observations of 6 new conventional heavy quarkonium states (4 $c\bar c$ and 2 $b\bar b$) have been made.

Exotic states such as tetraquarks and pentaquarks have lastly received considerable attention within the hadron physics community. Related with the first structures, the best known is the $X(3872)$, which was observed in 2003 as an extremely narrow peak in the $B^+ \to K^+ (\pi^+ \pi^- J/\psi)$ channel and at exactly the $\bar D^{0} D^{\ast0}$ threshold~\cite{Choi:2003ue, Aubert:2004ns}, and it is suspected to be a $c n \bar c \bar n$ ($n=u$ or $d$ quark) tetraquark state whose features resemble those of a molecule, but some experimental findings forbid to discard a more compact, diquark-antidiquark, component or even some $c\bar c$ trace in its wave function. On the other hand, there are no doubts of the tetraquark character of the $Z_c$'s~\cite{Ablikim:2013mio, Liu:2013dau} and $Z_b$'s~\cite{Belle:2011aa, Adachi:2012cx} states due to its non-zero charge. The most prominent examples of the second mentioned structures are the hidden-charm pentaquarks $P^+_c(4312)$, $P^+_c(4380)$, $P^+_c(4440)$ and $P^+_c(4457)$ reported in 2015 and 2019 by the LHCb collaboration in the $\Lambda_b^0$ decay, $\Lambda_b^0 \to J/\psi K^- p$~\cite{lhcb:2019pc, Aaij:2015tga}. 

The discussion about the nature of these exotic signals are carried out by various theoretical approaches. In particular, the three newly announced hidden-charm pentaquarks, $P^+_c(4312)$, $P^+_c(4440)$ and $P^+_c(4457)$ are favored to be molecular states of $\Sigma_c\bar{D}^*$ in, for instance, effective field theories~\cite{MZL190311560, JH190311872}, QCD sum rules~\cite{ZGW190502892}, phenomenological potential models~\cite{ZHG190400851, HH190400221, HM1904.09756, RZ190410285, MIE190411616, XZW190409891}, heavy quark spin symmetry formalisms~\cite{YS190400587, CWX190401296} and heavy hadron chiral perturbation theory~\cite{LM190504113}. Moreover, their photo-production~\cite{XC190406015, XYW190411706} and decay properties~\cite{CJX190400872} have been also discussed. As for the other types of pentaquarks, bound states of the $\bar{Q}qqqq$ system are not found within a constituent quark model~\cite{JMR190103578}. Using the same approach, several narrow double-heavy pentaquark states are found to be possible in the systematical investigations of Refs.~\cite{QSZ180104557, FG190304430, gy:2020dcp}. Moreover, within the one-boson-exchange model, possible triple-charm molecular pentaquarks $\Xi_{cc}D^{(*)}$ are suggested~\cite{FLW190101542}. In the tetraquark sector, double-heavy tetraquarks are studied using QCD sum rules~\cite{SSA190311975}, quark models~\cite{CEF190503877, gy:2020dht} and even lattice-regularized QCD computations~\cite{LL190404197}. Besides, theoretical techniques such as diffusion Monte Carlo~\cite{yb:2019plb}, Bethe-Salpeter equation~\cite{whge:2012plb}, QCD sum rules~\cite{zgwqqqq:2017epjc, wchxc:2017plb} and effective phenomenological models~\cite{mna:2018epjc, aeap:2018epjc, xc:2019epja, mslqfl:2019prd, gjw:2019arx} have recently contributed to the investigations of fully heavy tetraquarks $QQ\bar{Q}\bar{Q}$. Some reviews on both tetraquark and pentaquark systems can be found in Refs.~\cite{JV190209799, YRL190311976}.

Our QCD-inspired chiral quark model explained successfully the nature of the $P_c^+$ states in Ref.~\cite{Yang:2015bmv}, even before the last updated data reported by the LHCb collaboration~\cite{lhcb:2019pc}. Based on such fact, the  hidden-bottom~\cite{Yang:2018oqd} and double-charm pentaquarks~\cite{gy:2020dcp} were systematically investigated within the same theoretical framework, finding several either bound or resonance states. Reference~\cite{gy:2020dht} reported results on the double-heavy tetraquarks $QQ\bar{q}\bar{q}$ $(Q=c, b$ and $q=u, d)$, its natural extension should be the $QQ\bar{s}\bar{s}$ tetraquark sector with the hope of finding either bound or resonance states. In order to do so, we have recently established a complex scaling range formalism of the chiral quark model which allows us to determine (if exist) simultaneously scattering, resonance and bound states. We shall study herein the $QQ\bar{s}\bar{s}$ tetraquarks in the spin-parity channels $J^P=0^+$, $1^+$ and $2^+$, and in the isoscalar sector $I=0$. Another relevant feature of our study is that all configurations: meson-meson, diquark-antidiquark and K-type for four-body systems are considered; moreover, every possible color channel is taken into account, too. Finally, the Rayleigh-Ritz variational method is employed in dealing with the spatial wave functions of tetraquark states, which are expanded by means of the well-known Gaussian expansion method (GEM) of Ref.~\cite{Hiyama:2003cu}.

The present manuscript is arranged as follows. Section~\ref{sec:model} is devoted to briefly describe our theoretical approach which includes the complex-range formulation of the chiral quark model and the discussion of the $QQ\bar{s}\bar{s}$ wave-functions. Section~\ref{sec:results} is devoted to the analysis and discussion of the obtained results. The summary and some prospects are presented in Sec.~\ref{sec:summary}.

%%%%%%%%%%%%%%%%%%%%%%%%%%%%%%%%%%%%%%%%%%%%%%%%%%%%%%%%%%%%%%%%%%%%%%%%%%%%%%%%

\section{Theoretical framework}
\label{sec:model}

The complex scaling method (CSM) applied to our chiral quark model has been already explained in Refs.~\cite{gy:2020dht, gy:2020dcp}. The general form of the four-body complex Hamiltonian is given by
\begin{equation}
H(\theta) = \sum_{i=1}^{4}\left( m_i+\frac{\vec{p\,}^2_i}{2m_i}\right) - T_{\text{CM}} + \sum_{j>i=1}^{4} V(\vec{r}_{ij} e^{i\theta}) \,,
\label{eq:Hamiltonian}
\end{equation}
where the center-of-mass kinetic energy $T_{\text{CM}}$ is subtracted without loss of generality since we focus on the internal relative motions of quarks inside the multi-quark system. The interplay is of two-body potential which includes color-confining, $V_{\text{CON}}$, one-gluon exchange, $V_{\text{OGE}}$, and Goldstone-boson exchange, $V_{\chi}$, respectively,
\begin{equation}
\label{CQMV}
V(\vec{r}_{ij} e^{i\theta}) = V_{\text{CON}}(\vec{r}_{ij} e^{i\theta}) + V_{\text{OGE}}(\vec{r}_{ij} e^{i\theta}) + V_{\chi}(\vec{r}_{ij} e^{i\theta}) \,.
\end{equation}
In this work, we focus on the low-lying positive parity $QQ\bar{s}\bar{s}$ tetraquark states of $S$-wave, and hence only the central and spin-spin terms of the potentials shall be considered. 

By transforming the coordinates of relative motions between quarks as $\vec{r}_{ij} \rightarrow \vec{r}_{ij} e^{i\theta}$, the complex scaled Schr\"{o}dinger equation
\begin{equation}
\label{CSMSE}
\left[ H(\theta)-E(\theta) \right] \Psi_{JM}(\theta)=0 \,,
\end{equation}
is solved, giving eigenenergies that can be classified into three kinds of poles: bound, resonance and scattering ones, in a complex energy plane according to the so-called ABC theorem~\cite{JA22269, EB22280}. In particular, the resonance pole is independent of the rotated angle $\theta$, \emph{i.e.} it is fixed above the continuum cut line with a resonance's width $\Gamma=-2\,\text{Im}(E)$. The scattering state is just aligned along the cut line with a $2\theta$ rotated angle, whereas a bound state is always located on the real axis below its corresponding threshold.

The two-body potentials in Eq.~(\ref{CQMV}) mimic the most important features of QCD at low and intermediate energies. Firstly, color confinement should be encoded in the non-Abelian character of QCD. It has been demonstrated by lattice-QCD that multi-gluon exchanges produce an attractive linearly rising potential proportional to the distance between infinite-heavy quarks~\cite{Bali:2005fu}. However, the spontaneous creation of light-quark pairs from the QCD vacuum may give rise at the same scale to a breakup of the created color flux-tube~\cite{Bali:2005fu}. Therefore, the following expression when $\theta=0^\circ$ is used for the confinement potential:
\begin{equation}
V_{\text{CON}}(\vec{r}_{ij} e^{i\theta}\,)=\left[-a_{c}(1-e^{-\mu_{c}r_{ij} e^{i\theta}})+\Delta \right] 
(\vec{\lambda}_{i}^{c}\cdot\vec{\lambda}_{j}^{c}) \,,
\label{eq:conf}
\end{equation}
where $a_{c}$, $\mu_{c}$ and $\Delta$ are model parameters, and the SU(3) color Gell-Mann matrices are denoted as $\lambda^c$. One can see in Eq.~\eqref{eq:conf} that the potential is linear at short inter-quark distances with an effective confinement strength $\sigma = -a_{c} \, \mu_{c} \, (\vec{\lambda}^{c}_{i}\cdot \vec{\lambda}^{c}_{j})$, while $V_{\text{CON}}$ becomes constant $(\Delta-a_c)(\vec{\lambda}^{c}_{i}\cdot \vec{\lambda}^{c}_{j})$ at large distances. 

Secondly, the QCD's asymptotic freedom is expressed phenomenologically by the Fermi-Breit reduction of the one-gluon exchange interaction which, in the case of hadron systems with $\geq3$ quarks, consists on a Coulomb term supplemented by a chromomagnetic contact interaction given by
\begin{align}
&
V_{\text{OGE}}(\vec{r}_{ij} e^{i\theta}) = \frac{1}{4} \alpha_{s} (\vec{\lambda}_{i}^{c}\cdot
\vec{\lambda}_{j}^{c}) \Bigg[\frac{1}{r_{ij} e^{i\theta}} \nonumber \\ 
&
\hspace*{1.60cm} - \frac{1}{6m_{i}m_{j}} (\vec{\sigma}_{i}\cdot\vec{\sigma}_{j}) 
\frac{e^{-r_{ij} e^{i\theta}/r_{0}(\mu)}}{r_{ij} e^{i\theta} r_{0}^{2}(\mu)} \Bigg] \,,
\end{align}
where $m_{i}$ and $\vec{\sigma}$ are the quark mass and the Pauli matrices, respectively. The contact term of the central potential in complex range has been regularized as
\begin{equation}
\delta(\vec{r}_{ij} e^{i\theta})\sim\frac{1}{4\pi r_{0}^{2}}\frac{e^{-r_{ij} e^{i\theta}/r_{0}}}{r_{ij} e^{i\theta}} \,,
\end{equation}

The QCD-inspired effective scale-dependent strong coupling constant, $\alpha_s$, offers a consistent description of mesons and baryons from light to heavy quark sectors in wide energy range, and we use the frozen coupling constant of, for instance, Ref.~\cite{Segovia:2013wma}
\begin{equation}
\alpha_{s}(\mu_{ij})=\frac{\alpha_{0}}{\ln\left(\frac{\mu_{ij}^{2}+\mu_{0}^{2}}{\Lambda_{0}^{2}} \right)} \,,
\end{equation}
in which $\alpha_{0}$, $\mu_{0}$ and $\Lambda_{0}$ are parameters of the model.

Thirdly, the Goldstone-boson exchange interactions between light quarks, and constituent quark masses, appear because the breaking of chiral symmetry in a dynamical way. Therefore, the following two terms of the chiral potential must be taken into account between the $(\bar s\bar s)$-pair for $QQ\bar{s}\bar{s}$ tetraquarks:
\begin{align}
&
V_{\sigma}\left( \vec{r}_{ij} e^{i\theta} \right) = - \frac{g_{ch}^{2}}{4\pi}
\frac{\Lambda_{\sigma}^{2}}{\Lambda_{\sigma}^{2}-m_{\sigma}^{2}}m_{\sigma} \Bigg[
Y(m_{\sigma}r_{ij} e^{i\theta}) \nonumber \\
&
\hspace*{1.20cm} - \frac{\Lambda_{\sigma}}{m_{\sigma}}Y(\Lambda_{\sigma}r_{ij} e^{i\theta})
\Bigg] \,, \\
& 
V_{\eta}\left( \vec{r}_{ij} e^{i\theta} \right) = \frac{g_{ch}^{2}}{4\pi}
\frac{m_{\eta}^2}{12m_{i}m_{j}} \frac{\Lambda_{\eta}^{2}}{\Lambda_{\eta}^{2}-m_{
\eta}^{2}}m_{\eta} \Bigg[ Y(m_{\eta}r_{ij} e^{i\theta}) \nonumber \\
&
\hspace*{1.20cm} -\frac{\Lambda_{\eta}^{3}}{m_{\eta}^{3}
}Y(\Lambda_{\eta}r_{ij} e^{i\theta}) \Bigg] (\vec{\sigma}_{i}\cdot\vec{\sigma}_{j})
\Big[\cos\theta_{p} \left(\lambda_{i}^{8}\cdot\lambda_{j}^{8}
\right) \nonumber \\
&
\hspace*{1.20cm} -\sin\theta_{p} \Big] \,,
\end{align}
where $Y(x)=e^{-x}/x$ is the standard Yukawa function. The pion- and kaon-exchange interactions do not appear because no up- and down-quarks are considered herein. Furthermore, the physical $\eta$ meson is taken into account by introducing the angle $\theta_p$. The $\lambda^{a}$ are the SU(3) flavor Gell-Mann matrices. Taken from their experimental values, $m_{\pi}$, $m_{K}$ and $m_{\eta}$ are the masses of the SU(3) Goldstone bosons. The value of $m_{\sigma}$ is determined through the PCAC relation $m_{\sigma}^{2}\simeq m_{\pi}^{2}+4m_{u,d}^{2}$~\cite{Scadron:1982eg}. Finally, the chiral coupling constant, $g_{ch}$, is determined from the $\pi NN$ coupling constant through
\begin{equation}
\frac{g_{ch}^{2}}{4\pi}=\frac{9}{25}\frac{g_{\pi NN}^{2}}{4\pi} \frac{m_{u,d}^{2}}{m_{N}^2} \,,
\end{equation}
which assumes that flavor SU(3) is an exact symmetry only broken by the different mass of the strange quark.

The model parameters, which are listed in Table~\ref{tab:model}, have been fixed in advance reproducing hadron~\cite{Valcarce:1995dm, Vijande:2004he, Segovia:2008zza, Segovia:2008zz, Segovia:2009zz, Segovia:2011zza, Segovia:2015dia, Ortega:2016hde, Yang:2017xpp, Yang:2019lsg}, hadron-hadron~\cite{Fernandez:1993hx, Valcarce:1994nr, Ortega:2016mms, Ortega:2016pgg, Ortega:2017qmg} and multiquark~\cite{Vijande:2006jf, Yang:2015bmv, gy:2018dsichqm, Yang:2018oqd} phenomenology. Additionally, in order to help on our analysis of the $QQ\bar s\bar s$ tetraquarks in the following section, Table~\ref{MesonMass} lists the theoretical and experimental masses of the ground state and its first radial excitation (if available) for the $D^{(*)+}_s$ and $\bar{B}^{(*)}_s$ mesons. Besides, their mean-square radii are collected in Table~\ref{MesonMass}.

\begin{table}[!t]
\caption{\label{tab:model} Model parameters.}
\begin{ruledtabular}
\begin{tabular}{llr}
Quark masses     & $m_s$ (MeV) &  555 \\
                 & $m_c$ (MeV) & 1752 \\
                 & $m_b$ (MeV) & 5100 \\[2ex]
Goldstone bosons & $\Lambda_\sigma$ (fm$^{-1}$) & 4.20 \\
                 & $\Lambda_\eta$ (fm$^{-1}$)                & 5.20 \\
                 & $g^2_{ch}/(4\pi)$                         & 0.54 \\
                 & $\theta_P(^\circ)$                        & -15 \\[2ex]
Confinement      & $a_c$ (MeV)         & 430 \\
                 & $\mu_c$ (fm$^{-1})$ & 0.70 \\
                 & $\Delta$ (MeV)      & 181.10 \\[2ex]
                 & $\alpha_0$               & 2.118 \\
                 & $\Lambda_0~$ (fm$^{-1}$) & 0.113 \\
OGE              & $\mu_0~$ (MeV)        & 36.976 \\
                 & $\hat{r}_0~$ (MeV~fm) & 28.170 \\
\end{tabular}
\end{ruledtabular}
\end{table}

\begin{table}[!t]
\caption{\label{MesonMass} Theoretical and experimental masses of $D^{(*)+}_s$ and $B^{(*)}_s$ mesons; their mean-square radii are also shown.}
\begin{ruledtabular}
\begin{tabular}{lcccc}
Meson & $nL$ & $\surd{\langle r^2 \rangle}_{\text{The.}}$ (fm) & $M_{\text{The.}}$ (MeV) & $M_{\text{Exp.}}$ (MeV) \\
\hline
$D^+_s$ & $1S$ & 0.47 & 1989 & 1969 \\
        & $2S$ & 1.06 & 2703 & - \\[2ex]
$D^{*+}_s$ & $1S$ & 0.55 & 2116 & 2112 \\
           & $2S$ & 1.14 & 2767 & - \\[2ex]
$\bar{B}^0_s$ & $1S$ & 0.47 & 5355 & 5367 \\
              & $2S$ & 1.01 & 6017 & - \\[2ex]
$\bar{B}^*_s$ & $1S$ & 0.50 & 5400 & 5415 \\
              & $2S$ & 1.04 & 6042 & - \\ 
\end{tabular}
\end{ruledtabular}
\end{table}

\begin{figure}[ht]
\epsfxsize=3.4in \epsfbox{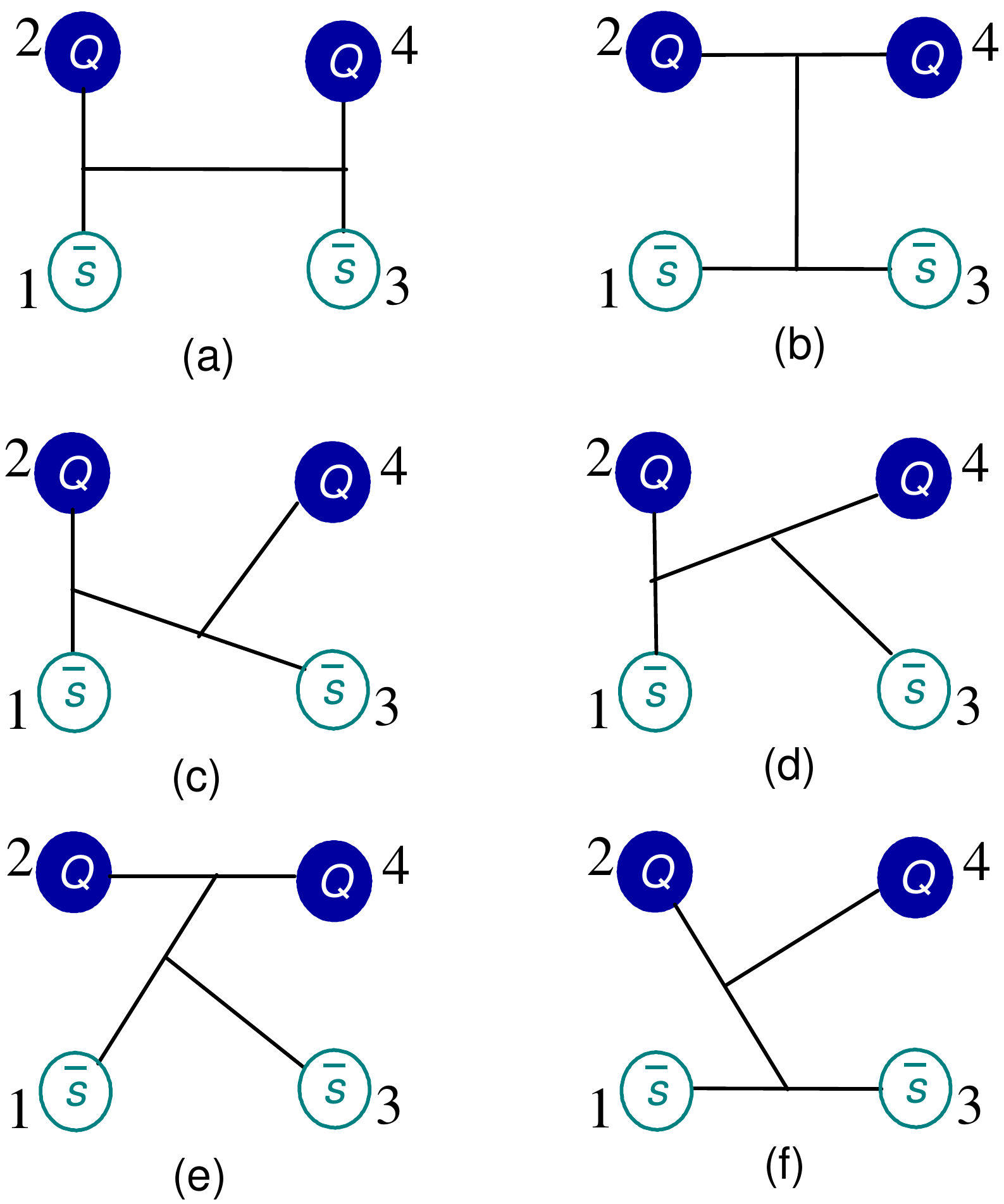}
\vspace*{-0.50cm}
\caption{Six types of configurations in $QQ\bar{s}\bar{s}$ $(Q=c,b)$ tetraquarks. \emph{Panel (a)} is the meson-meson configuration, \emph{panel (b)} is diquark-antidiquark one and the K-type structures are from \emph{panel (c)} to $(f)$.} \label{QQqq}
\end{figure}

Figure~\ref{QQqq} shows six kinds of configurations for double-heavy tetraquarks $QQ\bar{s}\bar{s}$ $(Q=c,b)$. In particular, Fig.~\ref{QQqq}(a) is the meson-meson (MM) structure, Fig.~\ref{QQqq}(b) is the diquark-antidiquark (DA) one, and the other K-type configurations are from panels (c) to (f). All of them, and their couplings, are considered in our investigation. However, for the purpose of solving a manageable 4-body problem, the K-type configurations are restricted to the case in which the two heavy quarks of $QQ\bar{s}\bar{s}$ tetraquarks are identical. It is important to note herein that just one configuration would be enough for the calculation, if all radial and orbital excited states were taken into account; however, this is obviously much less efficient and thus an economic way is to combine the different configurations in the ground state to perform the calculation.
% As a comparison with our previous work on $QQ\bar{q}\bar{q}$ $(q=u, d)$ tetraquarks~\cite{gy:2020dht}, herein much larger Hilbert spaces are considered with K-type configurations included.

Four fundamental degrees of freedom at the quark level: color, spin, flavor and space are generally accepted by QCD theory and the multiquark system's wave function is an internal product of color, spin, flavor and space terms. Firstly, concerning the color degree-of-freedom, plenty of color structures in multiquark system will be available with respect those of conventional hadrons ($q\bar{q}$ mesons and $qqq$ baryons). The colorless wave function of a 4-quark system in di-meson configuration, \emph{i.e.} as illustrated in Fig.~\ref{QQqq}(a), can be obtained by either a color-singlet or a hidden-color channel or both. However, this is not the unique way for the authors of Refs.~\cite{Harvey:1980rva, Vijande:2009kj}, who assert that it is enough to consider the color singlet channel when all possible excited states of a system are included.\footnote{After a comparison, a more economical way of computing through considering all the possible color structures and their coupling is preferred.} The $SU(3)_{\text{color}}$ wave functions of a color-singlet (two coupled color-singlet clusters, ${\bf 1}_c\otimes {\bf 1}_c$) and hidden-color (two coupled color-octet clusters, ${\bf 8}_c\otimes{\bf 8}_c$) channels are given by, respectively,
\begin{align}
\label{Color1}
\chi^c_1 &= \frac{1}{3}(\bar{r}r+\bar{g}g+\bar{b}b)\times (\bar{r}r+\bar{g}g+\bar{b}b) \,,
\end{align}
\begin{align}
\label{Color2}
\chi^c_2 &= \frac{\sqrt{2}}{12}(3\bar{b}r\bar{r}b+3\bar{g}r\bar{r}g+3\bar{b}g\bar{g}b+3\bar{g}b\bar{b}g+3\bar{r}g\bar{g}r
\nonumber\\
&+3\bar{r}b\bar{b}r+2\bar{r}r\bar{r}r+2\bar{g}g\bar{g}g+2\bar{b}b\bar{b}b-\bar{r}r\bar{g}g
\nonumber\\
&-\bar{g}g\bar{r}r-\bar{b}b\bar{g}g-\bar{b}b\bar{r}r-\bar{g}g\bar{b}b-\bar{r}r\bar{b}b) \,.
\end{align}
In addition, the color wave functions of the diquark-antidiquark structure shown in Fig.~\ref{QQqq}(b) are $\chi^c_3$ (color triplet-antitriplet clusters, ${\bf 3}_c\otimes {\bf \bar{3}}_c$) and $\chi^c_4$ (color sextet-antisextet clusters, ${\bf 6}_c\otimes{\bf \bar{6}}_c$), respectively:
\begin{align}
\label{Color3}
\chi^c_3 &= \frac{\sqrt{3}}{6}(\bar{r}r\bar{g}g-\bar{g}r\bar{r}g+\bar{g}g\bar{r}r-\bar{r}g\bar{g}r+\bar{r}r\bar{b}b
\nonumber\\
&-\bar{b}r\bar{r}b+\bar{b}b\bar{r}r-\bar{r}b\bar{b}r+\bar{g}g\bar{b}b-\bar{b}g\bar{g}b
\nonumber\\
&+\bar{b}b\bar{g}g-\bar{g}b\bar{b}g) \,,
\end{align}
\begin{align}
\label{Color4}
\chi^c_4 &= \frac{\sqrt{6}}{12}(2\bar{r}r\bar{r}r+2\bar{g}g\bar{g}g+2\bar{b}b\bar{b}b+\bar{r}r\bar{g}g+\bar{g}r\bar{r}g
\nonumber\\
&+\bar{g}g\bar{r}r+\bar{r}g\bar{g}r+\bar{r}r\bar{b}b+\bar{b}r\bar{r}b+\bar{b}b\bar{r}r
\nonumber\\
&+\bar{r}b\bar{b}r+\bar{g}g\bar{b}b+\bar{b}g\bar{g}b+\bar{b}b\bar{g}g+\bar{g}b\bar{b}g) \,.
\end{align}
Meanwhile, the colorless wave functions of the K-type structures shown in Fig.~\ref{QQqq}(c) to \ref{QQqq}(f) are obtained by following standard coupling algebra within the $SU(3)$ color group:\footnote{The group chain of K-type is obtained in sequence of quark number. Moreover, each quark and antiquark is represented, respectively, with [1] and [11] in the group theory.}
\begin{itemize}
\item	$K_1$-type of Fig.~\ref{QQqq}(c): \big[$C^{[21]}_{[1],[11]}C^{[221]}_{[21],[11]}C^{[222]}_{[221],[1]}$\big]$_5$; 

~~~~~~~~~~~~~~~~~~~~~~~~~~~~~\big[$C^{[111]}_{[1],[11]}C^{[221]}_{[111],[11]}C^{[222]}_{[221],[1]}$\big]$_6$;
\item	$K_2$-type of Fig.~\ref{QQqq}(d): \big[$C^{[111]}_{[1],[11]}C^{[211]}_{[111],[1]}C^{[222]}_{[211],[11]}$\big]$_7$;

~~~~~~~~~~~~~~~~~~~~~~~~~~~~~\big[$C^{[21]}_{[1],[11]}C^{[211]}_{[21],[1]}C^{[222]}_{[211],[11]}$\big]$_8$;
\item	$K_3$-type of Fig.~\ref{QQqq}(e): \big[$C^{[2]}_{[1],[1]}C^{[211]}_{[2],[11]}C^{[222]}_{[211],[11]}$\big]$_9$;

~~~~~~~~~~~~~~~~~~~~~~~~~~~~~\big[$C^{[11]}_{[1],[1]}C^{[211]}_{[11],[11]}C^{[222]}_{[211],[11]}$\big]$_{10}$;
\item	$K_4$-type of Fig.~\ref{QQqq}(f): \big[$C^{[22]}_{[11],[11]}C^{[221]}_{[22],[1]}C^{[222]}_{[221],[1]}$\big]$_{11}$;

~~~~~~~~~~~~~~~~~~~~~~~~~~~~~\big[$C^{[211]}_{[11],[11]}C^{[221]}_{[211],[1]}C^{[222]}_{[221],[1]}$\big]$_{12}$.
\end{itemize}
These group chains will generate the following K-type color wave functions whose subscripts correspond to those numbers above:
\begin{align}
\label{Color5}
\chi^c_5 &= \frac{1}{6\sqrt{2}}(\bar{r}r\bar{r}r+\bar{g}g\bar{g}g-2\bar{b}b\bar{b}b) \nonumber \\
&
+\frac{1}{2\sqrt{2}}(\bar{r}b\bar{b}r+\bar{r}g\bar{g}r+\bar{g}b\bar{b}g+\bar{g}r\bar{r}g+\bar{b}g\bar{g}b+\bar{b}r\bar{r}b) \nonumber \\
&
-\frac{1}{3\sqrt{2}}(\bar{g}g\bar{r}r+\bar{r}r\bar{g}g)+\frac{1}{6\sqrt{2}}(\bar{b}b\bar{r}r+\bar{b}b\bar{g}g+\bar{r}r\bar{b}b+\bar{g}g\bar{b}b) \,,
\end{align}
\begin{align}
\label{Color6}
\chi^c_6 &= \chi^c_1 \,,
\end{align}
\begin{align}
\label{Color7}
\chi^c_7 &= \chi^c_1 \,,
\end{align}
\begin{align}
\label{Color8}
\chi^c_8 &= \frac{1}{4}(1-\frac{1}{\sqrt{6}})\bar{r}r\bar{g}g-\frac{1}{4}(1+\frac{1}{\sqrt{6}})\bar{g}g\bar{g}g-\frac{1}{4\sqrt{3}}\bar{r}g\bar{g}r \nonumber \\
&
+\frac{1}{2\sqrt{2}}(\bar{r}b\bar{b}r+\bar{g}b\bar{b}g+\bar{b}g\bar{g}b+\bar{g}r\bar{r}g+\bar{b}r\bar{r}b) \nonumber \\
&
+\frac{1}{2\sqrt{6}}(\bar{r}r\bar{b}b-\bar{g}g\bar{b}b+\bar{b}b\bar{g}g+\bar{g}g\bar{r}r-\bar{b}b\bar{r}r) \,,
\end{align}
\begin{align}
\label{Color9}
\chi^c_9 &= \frac{1}{2\sqrt{6}}(\bar{r}b\bar{b}r+\bar{r}r\bar{b}b+\bar{g}b\bar{b}g+\bar{g}g\bar{b}b+\bar{r}g\bar{g}r+\bar{r}r\bar{g}g \nonumber\\
&
+\bar{b}b\bar{g}g+\bar{b}g\bar{g}b+\bar{g}g\bar{r}r+\bar{g}r\bar{r}g+\bar{b}b\bar{r}r+\bar{b}r\bar{r}b) \nonumber \\
&
+ \frac{1}{\sqrt{6}}(\bar{r}r\bar{r}r+\bar{g}g\bar{g}g+\bar{b}b\bar{b}b) \,,
\end{align}
\begin{align}
\label{Color10}
\chi^c_{10} &= \frac{1}{2\sqrt{3}}(\bar{r}b\bar{b}r-\bar{r}r\bar{b}b+\bar{g}b\bar{b}g-\bar{g}g\bar{b}b+\bar{r}g\bar{g}r-\bar{r}r\bar{g}g \nonumber \\
&
-\bar{b}b\bar{g}g+\bar{b}g\bar{g}b-\bar{g}g\bar{r}r+\bar{g}r\bar{r}g-\bar{b}b\bar{r}r+\bar{b}r\bar{r}b) \,,
\end{align}
\begin{align}
\label{Color11}
\chi^c_{11} &= \chi^c_9 \,,
\end{align}
\begin{align}
\label{Color12}
\chi^c_{12} &= -\chi^c_{10} \,,
\end{align}

As for the flavor degree-of-freedom, since the quark content of the tetraquark systems considered herein are two heavy quarks, $(Q=c,b)$, and two strange antiquarks, $\bar{s}$, only the isoscalar sector, $I=0$, will be discussed. The flavor wave functions denoted as $\chi^{fi}_{I, M_I}$, with the superscript $i=1,~2$ and $3$ referring to $cc\bar{s}\bar{s}$, $bb\bar{s}\bar{s}$ and $cb\bar{s}\bar{s}$ systems, can be written as
\begin{align}
&
\chi_{0,0}^{f1} = \bar{s}c\bar{s}c \,, \\
&
\chi_{0,0}^{f2} = \bar{s}b\bar{s}b \,, \\
&
\chi_{0,0}^{f3} = \bar{s}c\bar{s}b \,,
\end{align}
where, in this case, the third component of the isospin $M_I$ is equal to the value of total one $I$.

The total spin $S$ of tetraquark states ranges from $0$ to $2$. All of them shall be considered and, since there is not any spin-orbit potential, the third component $(M_S)$ can be set to be equal to the total one without loss of generality. Therefore, our spin wave functions $\chi^{\sigma_i}_{S, M_S}$ are given by
\begin{align}
\label{SWF1}
\chi_{0,0}^{\sigma_{l1}}(4) &= \chi^\sigma_{00}\chi^\sigma_{00} \,, \\
\chi_{0,0}^{\sigma_{l2}}(4) &= \frac{1}{\sqrt{3}}(\chi^\sigma_{11}\chi^\sigma_{1,-1}-\chi^\sigma_{10}\chi^\sigma_{10}+\chi^\sigma_{1,-1}\chi^\sigma_{11}) \,, \\
\chi_{0,0}^{\sigma_{l3}}(4) &= \frac{1}{\sqrt{2}} \Bigg( \Big( \sqrt{\frac{2}{3}} \chi^\sigma_{11} \chi^\sigma_{\frac{1}{2}, -\frac{1}{2}}-\sqrt{\frac{1}{3}} \chi^\sigma_{10} \chi^\sigma_{\frac{1}{2}, \frac{1}{2}} \Big) \chi^\sigma_{\frac{1}{2}, -\frac{1}{2}} \nonumber \\
&
-\Big( \sqrt{\frac{1}{3}} \chi^\sigma_{10} \chi^\sigma_{\frac{1}{2}, -\frac{1}{2}}-\sqrt{\frac{2}{3}} \chi^\sigma_{1, -1} \chi^\sigma_{\frac{1}{2}, \frac{1}{2}} \Big) \chi^\sigma_{\frac{1}{2}, \frac{1}{2}} \Bigg) \,, \\
\chi_{0,0}^{\sigma_{l4}}(4) &= \frac{1}{\sqrt{2}} \Big(\chi^\sigma_{00}\chi^\sigma_{\frac{1}{2}, \frac{1}{2}}\chi^\sigma_{\frac{1}{2}, -\frac{1}{2}}-\chi^\sigma_{00}\chi^\sigma_{\frac{1}{2}, -\frac{1}{2}}\chi^\sigma_{\frac{1}{2}, \frac{1}{2}} \Big) \,, \\
\chi_{1,1}^{\sigma_{m1}}(4) &= \chi^\sigma_{00}\chi^\sigma_{11} \,, \\ 
\chi_{1,1}^{\sigma_{m2}}(4) &= \chi^\sigma_{11}\chi^\sigma_{00} \,, \\
\chi_{1,1}^{\sigma_{m3}}(4) &= \frac{1}{\sqrt{2}} (\chi^\sigma_{11} \chi^\sigma_{10}-\chi^\sigma_{10} \chi^\sigma_{11}) \,, \\
\chi_{1,1}^{\sigma_{m4}}(4) &= \sqrt{\frac{3}{4}} \chi^\sigma_{11} \chi^\sigma_{\frac{1}{2}, \frac{1}{2}}\chi^\sigma_{\frac{1}{2}, -\frac{1}{2}}-\sqrt{\frac{1}{12}} \chi^\sigma_{11} \chi^\sigma_{\frac{1}{2}, -\frac{1}{2}} \chi^\sigma_{\frac{1}{2}, \frac{1}{2}} \nonumber \\ 
&
-\sqrt{\frac{1}{6}} \chi^\sigma_{10} \chi^\sigma_{\frac{1}{2}, \frac{1}{2}} \chi^\sigma_{\frac{1}{2}, \frac{1}{2}} \,, \\
\chi_{1,1}^{\sigma_{m5}}(4) &= \Bigg( \sqrt{\frac{2}{3}} \chi^\sigma_{11} \chi^\sigma_{\frac{1}{2}, -\frac{1}{2}}-\sqrt{\frac{1}{3}} \chi^\sigma_{10} \chi^\sigma_{\frac{1}{2}, \frac{1}{2}} \Bigg) \chi^\sigma_{\frac{1}{2}, \frac{1}{2}} \,, \\
\chi_{1,1}^{\sigma_{m6}}(4) &= \chi^\sigma_{00}\chi^\sigma_{\frac{1}{2}, \frac{1}{2}}\chi^\sigma_{\frac{1}{2}, \frac{1}{2}} \,, \\
\label{SWF2}
\chi_{2,2}^{\sigma_{1}}(4) &= \chi^\sigma_{11}\chi^\sigma_{11} \,.
\end{align}
The superscripts $l_1,\,\ldots\,,l_4$ and $m_1,\ldots,m_6$ are numbering the spin wave function for each configuration of tetraquark states, their specific values are shown in Table~\ref{SpinIndex}. Furthermore, these expressions are obtained by considering the coupling of two sub-cluster spin wave functions with SU(2) algebra, and the necessary bases read as
\begin{align}
\label{Spin}
\chi^\sigma_{11} &= \chi^\sigma_{\frac{1}{2}, \frac{1}{2}} \chi^\sigma_{\frac{1}{2}, \frac{1}{2}} \,, \\
\chi^\sigma_{1,-1} &= \chi^\sigma_{\frac{1}{2}, -\frac{1}{2}} \chi^\sigma_{\frac{1}{2}, -\frac{1}{2}} \,, \\
\chi^\sigma_{10} &= \frac{1}{\sqrt{2}}(\chi^\sigma_{\frac{1}{2}, \frac{1}{2}} \chi^\sigma_{\frac{1}{2}, -\frac{1}{2}}+\chi^\sigma_{\frac{1}{2}, -\frac{1}{2}} \chi^\sigma_{\frac{1}{2}, \frac{1}{2}})  \,, \\
\chi^\sigma_{00} &= \frac{1}{\sqrt{2}} \Big(\chi^\sigma_{\frac{1}{2}, \frac{1}{2}} \chi^\sigma_{\frac{1}{2}, -\frac{1}{2}}-\chi^\sigma_{\frac{1}{2}, -\frac{1}{2}} \chi^\sigma_{\frac{1}{2}, \frac{1}{2}} \Big) \,.
\end{align}
%and we could define $\chi^\sigma_{\frac{1}{2}, \frac{1}{2}}$ and $\chi^\sigma_{\frac{1}{2}, -\frac{1}{2}}$ as $\alpha$ and $\beta$ respectively.

\begin{table}[!t]
\caption{\label{SpinIndex} Index of spin wave function from Eq.~(\ref{SWF1}) to (\ref{SWF2}), their numbers are listed in the column of each configuration, respectively.}
\begin{ruledtabular}
\begin{tabular}{lcccccc}
    & Dimeson & Diquark-antidiquark & $K_1$ & $K_2$ & $K_3$ & $K_4$\\[2ex]
$l_1$ & 1   & 3 & & & & \\
$l_2$ & 2  & 4 & & & & \\
$l_3$ &   &  & 5 & 7 & 9 & 11 \\
$l_4$ &   &  & 6 & 8 & 10 & 12 \\[2ex]
$m_1$  & 1   & 4 & & & & \\
$m_2$ & 2  & 5 & & & & \\
$m_3$ & 3  & 6 & & & & \\
$m_4$ &   &  & 7 & 10 & 13 & 16 \\
$m_5$ &   &  & 8 & 11 & 14 & 17 \\
$m_6$ &   &  & 9 & 12 & 15 & 18
\end{tabular}
\end{ruledtabular}
\end{table}

Among the different methods to solve the Schr\"odinger-like 4-body bound state equation, we use the Rayleigh-Ritz variational principle which is one of the most extended tools to solve eigenvalue problems because its simplicity and flexibility. Meanwhile, the choice of basis to expand the wave function solution is of great importance. Within hte CRM, the spatial wave function can be written as follows
\begin{equation}
\label{eq:WFexp}
\psi_{LM_L}(\theta)= \left[ \left[ \phi_{n_1l_1}(\vec{\rho}e^{i\theta}\,) \phi_{n_2l_2}(\vec{\lambda}e^{i\theta}\,)\right]_{l} \phi_{n_3l_3}(\vec{R}e^{i\theta}\,) \right]_{L M_L} \,,
\end{equation}
where the internal Jacobi coordinates for the meson-meson configuration (Fig.~\ref{QQqq}(a)) are defined as
\begin{align}
\vec{\rho} &= \vec{x}_1-\vec{x}_2 \,, \\
\vec{\lambda} &= \vec{x}_3 - \vec{x}_4 \,, \\
\vec{R} &= \frac{m_1 \vec{x}_1 + m_2 \vec{x}_2}{m_1+m_2}- \frac{m_3 \vec{x}_3 + m_4 \vec{x}_4}{m_3+m_4} \,,
\end{align}
and for the diquark-antdiquark one (Fig.~\ref{QQqq}(b)) are
\begin{align}
\vec{\rho} &= \vec{x}_1-\vec{x}_3 \,, \\
\vec{\lambda} &= \vec{x}_2 - \vec{x}_4 \,, \\
\vec{R} &= \frac{m_1 \vec{x}_1 + m_3 \vec{x}_3}{m_1+m_3}- \frac{m_2 \vec{x}_2 + m_4 \vec{x}_4}{m_2+m_4} \,.
\end{align}
The Jacobi coordinates for the remaining K-type configurations shown in Fig.~\ref{QQqq}, panels (c) to (f), are ($i, j, k, l$ are according to the definitions of each configuration in Fig.~\ref{QQqq}):
\begin{align}
\vec{\rho} &= \vec{x}_i-\vec{x}_j \,, \\
\vec{\lambda} &= \vec{x}_k- \frac{m_i \vec{x}_i + m_j \vec{x}_j}{m_i+m_j} \,, \\
\vec{R} &= \vec{x}_l- \frac{m_i \vec{x}_i + m_j \vec{x}_j+m_k \vec{x}_k}{m_i+m_j+m_k} \,.
\end{align}
Obviously, the center-of-mass kinetic term $T_{\text{CM}}$ can be completely eliminated for a non-relativistic system when using these sets of coordinates.

A very efficient method to solve the bound-state problem of a few-body system is the Gaussian expansion method (GEM)~\cite{Hiyama:2003cu}, which has been successfully applied by us in other multiquark systems~\cite{Yang:2015bmv, Yang:2018oqd, gy:2020dcp, gy:2020dht}. The Gaussian basis in each relative coordinate is taken with geometric progression in the size parameter.\footnote{The details on Gaussian parameters and how they are fixed can be found in Ref.~\cite{Yang:2015bmv}.} Therefore, the form of the orbital wave functions, $\phi$'s, in Eq.~\eqref{eq:WFexp} is 
\begin{align}
&
\phi_{nlm}(\vec{r}e^{i\theta}\,) = N_{nl} (re^{i\theta})^{l} e^{-\nu_{n} (re^{i\theta})^2} Y_{lm}(\hat{r}) \,.
%&
%= N_{nl} \lim_{\varepsilon\to 0} \frac{1}{(\nu_{n}\varepsilon)^l} \sum_{k=1}^{k_{\rm
%max}} C_{lm,k} e^{-\nu_{n}(\vec{r}-\varepsilon \vec{D}_{lm,k})^{2}} \,.
\end{align}
As one can see, the Jacobi coordinates are all transformed with a common scaling angle $\theta$ in the complex scaling method. In this way, both bound states and resonances can be described simultaneously within one scheme. Moreover, only $S$-wave state of double-heavy tetraquarks are investigated in this work and thus no laborious Racah algebra is needed during matrix elements calculation.
%because the value of spherical harmonic function is a constant when $l=0$, $i. e.$ $Y_{00}=\sqrt{1/4\pi}$.

Finally, in order to fulfill the Pauli principle, the complete wave-function is written as
\begin{equation}
\label{TPs}
\Psi_{JM_J,I,i,j,k}(\theta)={\cal A} \left[ \left[ \psi_{L}(\theta) \chi^{\sigma_i}_{S}(4) \right]_{JM_J} \chi^{f_j}_I \chi^{c}_k \right] \,,
\end{equation}
where $\cal{A}$ is the antisymmetry operator of $QQ\bar{s}\bar{s}$ tetraquarks when considering interchange between identical particles ($\bar{s}\bar{s},~cc$ and $bb$). This is necessary because the complete wave function of the 4-quark system is constructed from two sub-clusters: meson-meson, diquark-antidiquark and K-type structures. In particular, when the two heavy quarks are of the same flavor ($QQ=cc$ or $bb$), the operator $\cal{A}$ with the quark arrangements $\bar{s}Q\bar{s}Q$ is defined as
\begin{equation}
{\cal{A}} = 1-(13)-(24)+(13)(24) \,.
\end{equation}
However, due to the fact that $c$- and $b$-quarks are distinguishable particles, the operator $\cal{A}$ consists only on two terms for the $\bar{s}c\bar{s}b$ system, and read as
\begin{equation}
{\cal{A}} = 1-(13) \,.
\end{equation}

%%%%%%%%%%%%%%%%%%%%%%%%%%%%%%%%%%%%%%%%%%%%%%%%%%%%%%%%%%%%%%%%%%%%%%%%%%%%%%%%

\begin{table}[!t]
\caption{\label{GDD1} All possible channels for $IJ^P=00^+$ $cc\bar{s}\bar{s}$ and $bb\bar{s}\bar{s}$ tetraquark systems. The second column shows the necessary basis combination in spin ($\chi_J^{\sigma_i}$), flavor ($\chi_I^{f_j}$) and color ($\chi_k^c$) degrees of freedom. Particularly, the flavor indices ($j$) 1 and 2 are of $cc\bar{s}\bar{s}$ and $bb\bar{s}\bar{s}$, respectively. The superscript 1 and 8 stands for the color-singlet and hidden-color configurations of physical channels.}
\begin{ruledtabular}
\begin{tabular}{ccc}
~~Index & $\chi_J^{\sigma_i}$;~$\chi_I^{f_j}$;~$\chi_k^c$ & Channel~~\\
              &$[i; ~j; ~k]$&  \\[2ex]
1  & $[1; ~1\,(2); ~1]$   & $(D^+_s D^+_s)^1$; $(\bar{B}^0_s \bar{B}^0_s)^1$ \\
2 & $[2; ~1\,(2); ~1]$  & $(D^{*+}_s D^{*+}_s)^1$; $(\bar{B}^*_s \bar{B}^*_s)^1$ \\
3 & $[1; ~1\,(2); ~2]$  & $(D^+_s D^+_s)^8$; $(\bar{B}^0_s \bar{B}^0_s)^8$ \\
4 & $[2; ~1\,(2); ~2]$  & $(D^{*+}_s D^{*+}_s)^8$; $(\bar{B}^*_s \bar{B}^*_s)^8$ \\
5 & $[3; ~1\,(2); ~4]$  & $(cc)(\bar{s}\bar{s})$; $(bb)(\bar{s}\bar{s})$ \\
6 & $[4; ~1\,(2); ~3]$  & $(cc)^*(\bar{s}\bar{s})^*$; $(bb)^*(\bar{s}\bar{s})^*$ \\
7 & $[5; ~1\,(2); ~5]$  & $K_1$ \\
8 & $[5; ~1\,(2); ~6]$   & $K_1$ \\
9 & $[6; ~1\,(2); ~5]$  & $K_1$ \\
10 & $[6; ~1\,(2); ~6]$  & $K_1$ \\
11 & $[7; ~1\,(2); ~7]$  & $K_2$ \\
12 & $[7; ~1\,(2); ~8]$  & $K_2$ \\
13 & $[8; ~1\,(2); ~7]$  & $K_2$ \\
14 & $[8; ~1\,(2); ~8]$  & $K_2$ \\
15 & $[9; ~1\,(2); ~10]$   & $K_3$ \\
16 & $[10; ~1\,(2); ~9]$  & $K_3$ \\
17 & $[11; ~1\,(2); ~12]$  & $K_4$ \\
18 & $[12; ~1\,(2); ~11]$  & $K_4$
\end{tabular}
\end{ruledtabular}
\end{table}

\begin{table}[!t]
\caption{\label{GDD2} All possible channels for $IJ^P=01^+$ $cc\bar{s}\bar{s}$ and $bb\bar{s}\bar{s}$ tetraquark systems. The second column shows the necessary basis combination in spin ($\chi_J^{\sigma_i}$), flavor ($\chi_I^{f_j}$) and color ($\chi_k^c$) degrees of freedom. Particularly, the flavor indices ($j$) 1 and 2 are of $cc\bar{s}\bar{s}$ and $bb\bar{s}\bar{s}$, respectively. The superscript 1 and 8 stands for the color-singlet and hidden-color configurations of physical channels.}
\begin{ruledtabular}
\begin{tabular}{ccc}
~~Index & $\chi_J^{\sigma_i}$;~$\chi_I^{f_j}$;~$\chi_k^c$ & Channel~~\\
              &$[i; ~j; ~k]$&  \\[2ex]
1  & $[1; ~1\,(2); ~1]$   & $(D^+_s D^{*+}_s)^1$; $(\bar{B}^0_s \bar{B}^*_s)^1$ \\
2 & $[3; ~1\,(2); ~1]$  & $(D^{*+}_s D^{*+}_s)^1$; $(\bar{B}^*_s \bar{B}^*_s)^1$ \\
3 & $[1; ~1\,(2); ~2]$  & $(D^+_s D^{*+}_s)^8$; $(\bar{B}^0_s \bar{B}^*_s)^8$\\
4 & $[3; ~1\,(2); ~2]$  & $(D^{*+}_s D^{*+}_s)^8$; $(\bar{B}^*_s \bar{B}^*_s)^8$\\
5 & $[6; ~1\,(2); ~3]$  & $(cc)^*(\bar{s}\bar{s})^*$; $(bb)^*(\bar{s}\bar{s})^*$ \\
6 & $[7; ~1\,(2); ~5]$  & $K_1$ \\
7 & $[8; ~1\,(2); ~5]$  & $K_1$ \\
8 & $[9; ~1\,(2); ~5]$   & $K_1$ \\
9 & $[7; ~1\,(2); ~6]$  & $K_1$ \\
10 & $[8; ~1\,(2); ~6]$  & $K_1$ \\
11 & $[9; ~1\,(2); ~6]$  & $K_1$ \\
12 & $[10; ~1\,(2); ~7]$  & $K_2$ \\
13 & $[11; ~1\,(2); ~7]$  & $K_2$ \\
14 & $[12; ~1\,(2); ~7]$  & $K_2$ \\
15 & $[10; ~1\,(2); ~8]$   & $K_2$ \\
16 & $[11; ~1\,(2); ~8]$  & $K_2$ \\
17 & $[12; ~1\,(2); ~8]$  & $K_2$ \\
18 & $[13; ~1\,(2); ~10]$  & $K_3$ \\
19 & $[14; ~1\,(2); ~10]$  & $K_3$ \\
20 & $[15; ~1\,(2); ~9]$  & $K_3$ \\
21 & $[16; ~1\,(2); ~12]$  & $K_4$ \\
22 & $[17; ~1\,(2); ~12]$  & $K_4$ \\
23 & $[18; ~1\,(2); ~11]$  & $K_4$
\end{tabular}
\end{ruledtabular}
\end{table}

\begin{table}[!t]
\caption{\label{GDD3} All possible channels for $IJ^P=02^+$ $cc\bar{s}\bar{s}$ and $bb\bar{s}\bar{s}$ tetraquark systems. The second column shows the necessary basis combination in spin ($\chi_J^{\sigma_i}$), flavor ($\chi_I^{f_j}$) and color ($\chi_k^c$) degrees of freedom. Particularly, the flavor indices ($j$) 1 and 2 are of $cc\bar{s}\bar{s}$ and $bb\bar{s}\bar{s}$, respectively. The superscript 1 and 8 stands for the color-singlet and hidden-color configurations of physical channels.}
\begin{ruledtabular}
\begin{tabular}{ccc}
~~Index & $\chi_J^{\sigma_i}$;~$\chi_I^{f_j}$;~$\chi_k^c$ & Channel~~\\
              &$[i; ~j; ~k]$&  \\[2ex]
1  & $[1; ~1\,(2); ~1]$   & $(D^{*+}_s D^{*+}_s)^1$; $(\bar{B}^*_s \bar{B}^*_s)^1$\\
2 & $[1; ~1\,(2); ~2]$  & $(D^{*+}_s D^{*+}_s)^8$; $(\bar{B}^*_s \bar{B}^*_s)^8$ \\
3 & $[1; ~1\,(2); ~3]$  & $(cc)^*(\bar{s}\bar{s})^*$; $(bb)^*(\bar{s}\bar{s})^*$ \\
4 & $[1; ~1\,(2); ~5]$  & $K_1$ \\
5 & $[1; ~1\,(2); ~6]$  & $K_1$ \\
6 & $[1; ~1\,(2); ~7]$  & $K_2$ \\
7 & $[1; ~1\,(2); ~8]$  & $K_2$ \\
8 & $[1; ~1\,(2); ~10]$   & $K_3$ \\
9 & $[1; ~1\,(2); ~12]$  & $K_4$
\end{tabular}
\end{ruledtabular}
\end{table}

\begin{table}[!t]
\caption{\label{GDD4} All possible channels for $IJ^P=00^+$ $cb\bar{s}\bar{s}$ tetraquark systems. The second column shows the necessary basis combination in spin ($\chi_J^{\sigma_i}$), flavor ($\chi_I^{f_j}$) and color ($\chi_k^c$) degrees of freedom. The superscript 1 and 8 stands for the color-singlet and hidden-color configurations of physical channels.}
\begin{ruledtabular}
\begin{tabular}{ccc}
~~Index & $\chi_J^{\sigma_i}$;~$\chi_I^{f_j}$;~$\chi_k^c$ & Channel~~\\
              &$[i; ~j; ~k]$&  \\[2ex]
1  & $[1; ~3; ~1]$   & $(D^+_s \bar{B}^0_s)^1$ \\
2 & $[2; ~3; ~1]$  & $(D^{*+}_s \bar{B}^*_s)^1$ \\
3 & $[1; ~3; ~2]$  & $(D^+_s \bar{B}^0_s)^8$\\
4 & $[2; ~3; ~2]$  & $(D^{*+}_s \bar{B}^*_s)^8$\\
5 & $[3; ~3; ~4]$  & $(cb)(\bar{s}\bar{s})$ \\
6 & $[4; ~3; ~3]$  & $(cb)^*(\bar{s}\bar{s})^*$ \\
7 & $[5; ~3; ~5]$  & $K_1$ \\
8 & $[5; ~3; ~6]$   & $K_1$ \\
9 & $[6; ~3; ~5]$  & $K_1$ \\
10 & $[6; ~3; ~6]$  & $K_1$ \\
11 & $[7; ~3; ~7]$  & $K_2$ \\
12 & $[7; ~3; ~8]$  & $K_2$ \\
13 & $[8; ~3; ~7]$  & $K_2$ \\
14 & $[8; ~3; ~8]$  & $K_2$ \\
15 & $[9; ~3; ~9]$   & $K_3$ \\
16 & $[9; ~3; ~10]$  & $K_3$ \\
17 & $[10; ~3; ~9]$  & $K_3$ \\
18 & $[10; ~3; ~10]$  & $K_3$ \\
19 & $[11; ~3; ~12]$  & $K_4$ \\
20 & $[12; ~3; ~11]$  & $K_4$
\end{tabular}
\end{ruledtabular}
\end{table}

\begin{table}[!t]
\caption{\label{GDD5} All possible channels for $IJ^P=01^+$ $cb\bar{s}\bar{s}$ tetraquark systems. The second column shows the necessary basis combination in spin ($\chi_J^{\sigma_i}$), flavor ($\chi_I^{f_j}$) and color ($\chi_k^c$) degrees of freedom. The superscript 1 and 8 stands for the color-singlet and hidden-color configurations of physical channels.}
\begin{ruledtabular}
\begin{tabular}{ccc}
~~Index & $\chi_J^{\sigma_i}$;~$\chi_I^{f_j}$;~$\chi_k^c$ & Channel~~\\
              &$[i; ~j; ~k]$&  \\[2ex]
1  & $[1; ~3; ~1]$   & $(D^+_s \bar{B}^*_s)^1$ \\
2 & $[2; ~3; ~1]$  & $(D^{*+}_s \bar{B}^0_s)^1$ \\
3 & $[3; ~3; ~1]$  & $(D^{*+}_s \bar{B}^*_s)^1$\\
4 & $[1; ~3; ~2]$  & $(D^+_s \bar{B}^*_s)^8$\\
5 & $[2; ~3; ~2]$  & $(D^{*+}_s \bar{B}^0_s)^8$ \\
6 & $[3; ~3; ~2]$  & $(D^{*+}_s \bar{B}^*_s)^8$ \\
7 & $[6; ~3; ~3]$  & $(cb)(\bar{s}\bar{s})^*$ \\
8 & $[5; ~3; ~3]$   & $(cb)^*(\bar{s}\bar{s})$ \\
9 & $[4; ~3; ~4]$  & $(cb)^*(\bar{s}\bar{s})^*$ \\
10 & $[7; ~3; ~5]$  & $K_1$ \\
11 & $[8; ~3; ~5]$  & $K_1$ \\
12 & $[9; ~3; ~5]$  & $K_1$ \\
13 & $[7; ~3; ~6]$  & $K_1$ \\
14 & $[8; ~3; ~6]$  & $K_1$ \\
15 & $[9; ~3; ~6]$   & $K_1$ \\
16 & $[10; ~3; ~7]$  & $K_2$ \\
17 & $[11; ~3; ~7]$  & $K_2$ \\
18 & $[12; ~3; ~7]$  & $K_2$ \\
19 & $[10; ~3; ~8]$  & $K_2$ \\
20 & $[11; ~3; ~8]$  & $K_2$ \\
21 & $[12; ~3; ~8]$  & $K_2$ \\
22 & $[13; ~3; ~10]$  & $K_3$ \\
23 & $[14; ~3; ~10]$  & $K_3$ \\
24 & $[15; ~3; ~10]$  & $K_3$ \\
25 & $[13; ~3; ~9]$   & $K_3$ \\
26 & $[14; ~3; ~9]$  & $K_3$ \\
27 & $[15; ~3; ~9]$  & $K_3$ \\
28 & $[16; ~3; ~12]$  & $K_4$ \\
29 & $[17; ~3; ~12]$  & $K_4$ \\
30 & $[18; ~3; ~11]$  & $K_4$
\end{tabular}
\end{ruledtabular}
\end{table}

\begin{table}[!t]
\caption{\label{GDD6} All possible channels for $IJ^P=02^+$ $cb\bar{s}\bar{s}$ tetraquark systems. The second column shows the necessary basis combination in spin ($\chi_J^{\sigma_i}$), flavor ($\chi_I^{f_j}$) and color ($\chi_k^c$) degrees of freedom. The superscript 1 and 8 stands for the color-singlet and hidden-color configurations of physical channels.}
\begin{ruledtabular}
\begin{tabular}{ccc}
~~Index & $\chi_J^{\sigma_i}$;~$\chi_I^{f_j}$;~$\chi_k^c$ & Channel~~\\
              &$[i; ~j; ~k]$&  \\[2ex]
1  & $[1; ~3; ~1]$   & $(D^{*+}_s \bar{B}^*_s)^1$ \\
2 & $[1; ~3; ~2]$  & $(D^{*+}_s \bar{B}^*_s)^8$ \\
3 & $[1; ~3; ~3]$  & $(cb)^*(\bar{s}\bar{s})^*$\\
4 & $[1; ~3; ~5]$  & $K_1$\\
5 & $[1; ~3; ~6]$  & $K_1$ \\
6 & $[1; ~3; ~7]$  & $K_2$ \\
7 & $[1; ~3; ~8]$  & $K_2$ \\
8 & $[1; ~3; ~9]$   & $K_3$ \\
9 & $[1; ~3; ~10]$  & $K_3$ \\
10 & $[1; ~3; ~12]$  & $K_4$
\end{tabular}
\end{ruledtabular}
\end{table}

\section{Results}
\label{sec:results}

The low-lying $S$-wave states of $QQ\bar{s}\bar{s}$ $(Q=c,b)$ tetraquarks are systematically investigated herein. The parity for different $QQ\bar{s}\bar{s}$ tetraquarks is positive under our assumption that the angular momenta $l_1$, $l_2$, $l_3$, which appear in Eq.~\eqref{eq:WFexp}, are all $0$. Accordingly, the total angular momentum, $J$, coincides with the total spin, $S$, and can take values $0$, $1$ and $2$. Note, too, the value of isospin can only be $0$ for the $QQ\bar{s}\bar{s}$ system. For $cc\bar{s}\bar{s}$, $bb\bar{s}\bar{s}$ and $cb\bar{s}\bar{s}$ systems, all possible meson-meson, diquark-antidiquark and K-type channels for each $IJ^P$ quantum numbers are listed in Table~\ref{GDD1}, \ref{GDD2}, \ref{GDD3}, \ref{GDD4}, \ref{GDD5} and \ref{GDD6}, respectively. The second column shows the necessary basis combination in spin $(\chi^{\sigma_i}_J)$, flavor $(\chi^{f_j}_I)$, and color $(\chi^c_k)$ degrees-of-freedom. The physical channels with color-singlet (labeled with the superindex $1$), hidden-color (labeled with the superindex $8$), diquark-antidiquark (labeled with $(QQ)(\bar{s}\bar{s})$) and K-type (labeled from $K_1$ to $K_4$) configurations are listed in the third column. %All of those possible channels in a quantum state are numbered in the first column.

Tables ranging from~\ref{GresultCC1} to~\ref{GresultCCT} summarize our calculated results (mass and width) of the lowest-lying $QQ\bar{s}\bar{s}$ tetraquark states and possible resonances. In particular, results of $cc\bar{s}\bar{s}$ tetraquarks with $I(J^P)=0(0^+)$, $0(1^+)$ and $0(2^+)$ are listed in Tables~\ref{GresultCC1}, \ref{GresultCC2} and \ref{GresultCC3}; those of $bb\bar{s}\bar{s}$ tetraquarks are shown in Tables~\ref{GresultCC4}, \ref{GresultCC5} and \ref{GresultCC6}; and Tables~\ref{GresultCC7}, \ref{GresultCC8} and \ref{GresultCC9} collect the $cb\bar{s}\bar{s}$ cases. In these tables, the first column lists the physical channel of meson-meson, diquark-antidiquark and K-type (if it fulfills Pauli principle), and the experimental value of the noninteracting meson-meson threshold is also indicated in parenthesis; the second column signals the discussed channel, \emph{e.g.} color-singlet (S), hidden-color (H), etc.; the third column shows the theoretical mass $(M)$ of each single channel; and the fourth column shows a coupled calculation result for one certain configuration. Moreover, the complete coupled channels results for each quantum state are shown at the bottom of each table. Besides, Table~\ref{GresultCCT} summarizes the obtained resonance states of $QQ\bar{s}\bar{s}$ tetraquarks in the complete coupled-channels calculation. 

\begin{figure}[ht]
\epsfxsize=3.85in \epsfbox{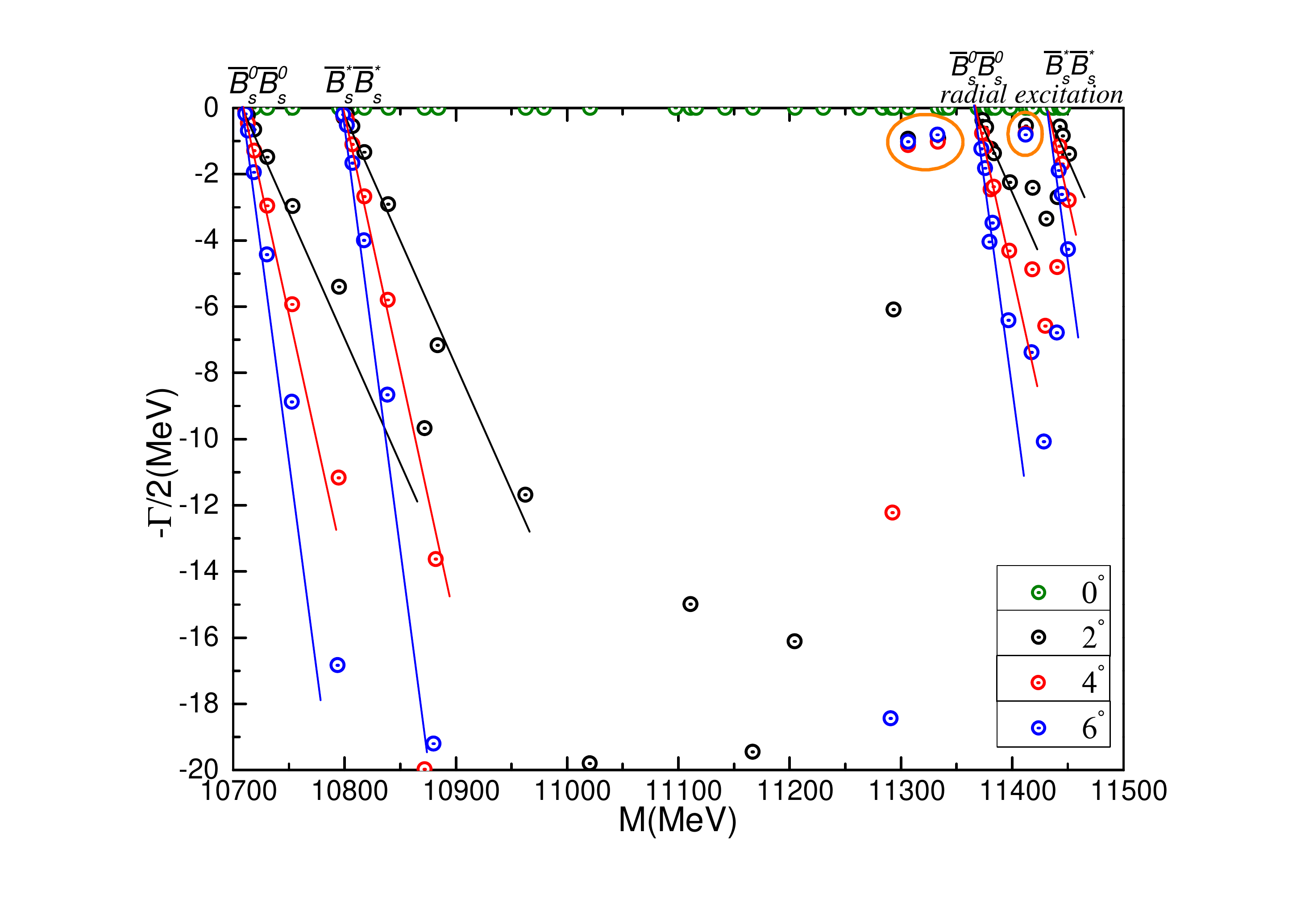}
\vspace*{-1.10cm}
\caption{Complex energies of $bb\bar{s}\bar{s}$ tetraquarks with $IJ^P=00^+$ in the coupled channels calculation, $\theta$ varying from $0^\circ$ to $6^\circ$ .} \label{PP4}
\end{figure}

\begin{figure}[ht]
\epsfxsize=3.85in \epsfbox{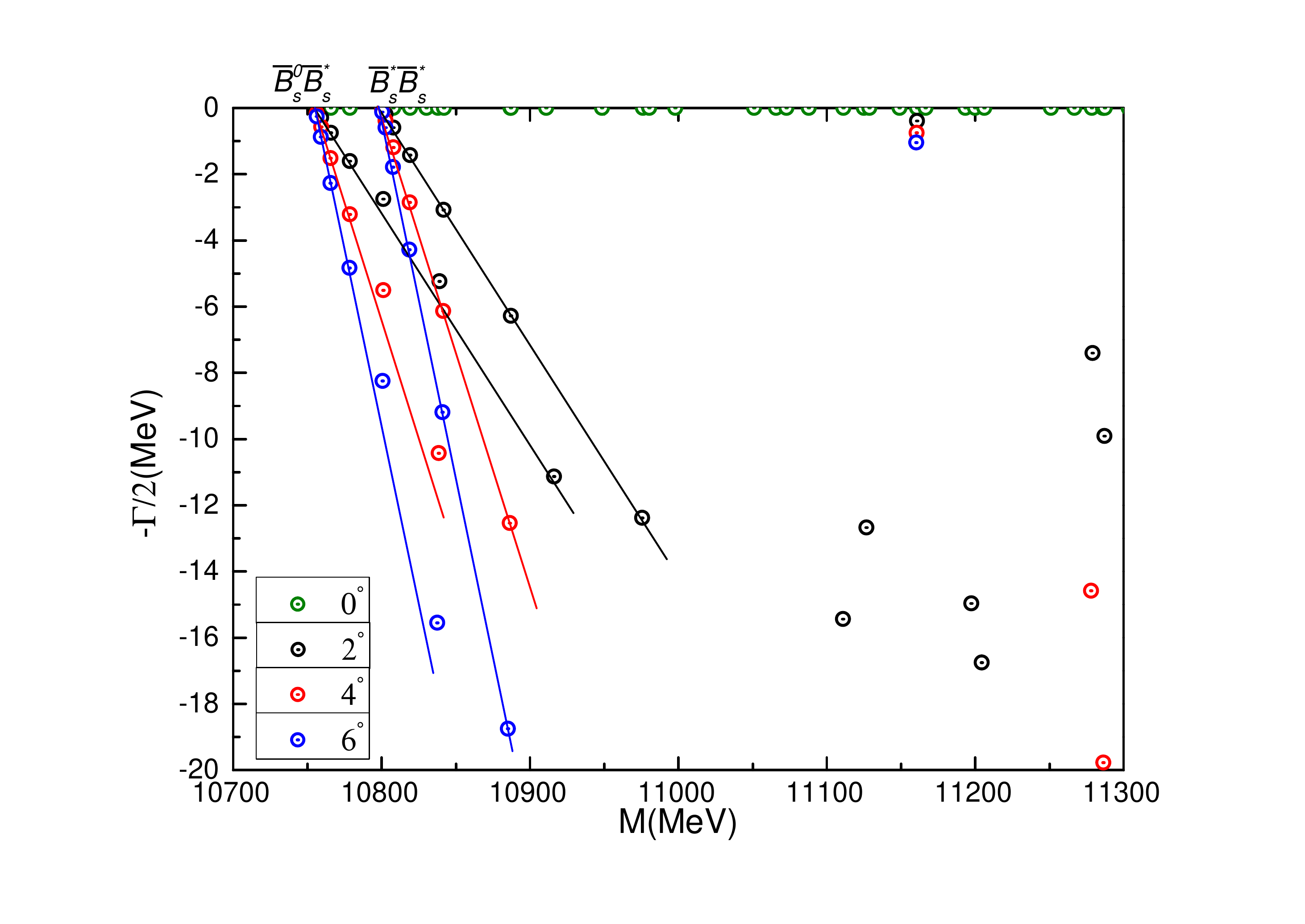}
\vspace*{-1.10cm}
\caption{Complex energies of $bb\bar{s}\bar{s}$ tetraquarks with $IJ^P=01^+$ in the coupled channels calculation, $\theta$ varying from $0^\circ$ to $6^\circ$ .} \label{PP5}
\end{figure}

\begin{figure}[ht]
\epsfxsize=3.85in \epsfbox{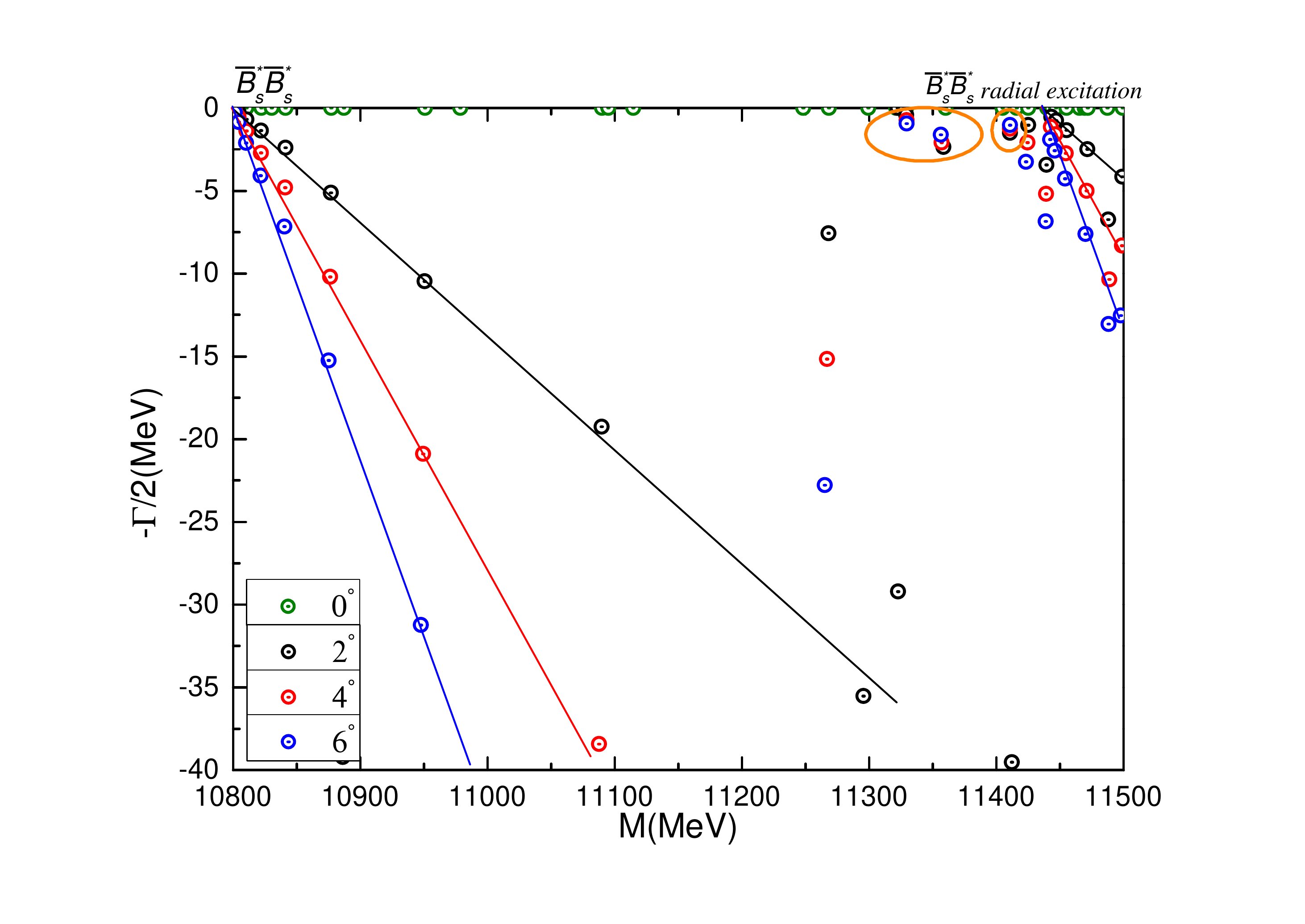}
\vspace*{-1.10cm}
\caption{Complex energies of $bb\bar{s}\bar{s}$ tetraquarks with $IJ^P=02^+$ in the coupled channels calculation, $\theta$ varying from $0^\circ$ to $6^\circ$ .} \label{PP6}
\end{figure}

Figures~\ref{PP1} to~\ref{PP9} depict the distribution of complex energies of the $QQ\bar{s}\bar{s}$ tetraquarks in the complete coupled-channels calculation. The $x$-axis is the real part of the complex energy $E$, which stands for the mass of tetraquark states, and the $y$-axis is the imaginary part of $E$, which is related to the width through $\Gamma=-2\,\text{Im}(E)$. In the figures, some orange circles appear surrounding resonance candidates. They are usually $\sim$ 0.6 GeV above their respective non-interacting meson-meson thresholds and $\sim$ 0.2 GeV around their first radial excitation states; moreover, looking at the details, we shall conclude that most of these observed resonances can be identified with a hadronic molecular nature.
% Looking at the sizes of the conventional $Q\bar{s}$ mesons listed in Table~\ref{MesonMass} and the typical size, $\sim2\,\text{fm}$, of the only well established hadronic molecule: the deuteron, most of these observed resonances can be identified of molecular nature.

Now let us proceed to describe in detail our theoretical findings for each sector of $QQ\bar{s}\bar{s}$ tetraquarks.

\begin{table}[!t]
\caption{\label{GresultCC1} The lowest-lying eigen-energies of $cc\bar{s}\bar{s}$ tetraquarks with $IJ^P=00^+$ in the real range calculation. The first column shows the allowed channels and, in the parenthesis, the noninteracting meson-meson threshold value of experiment. Color-singlet (S), hidden-color (H) along with other configurations are indexed in the second column respectively, the third and fourth columns refer to the theoretical mass of each channels and their couplings. (unit: MeV)}
\begin{ruledtabular}
\begin{tabular}{lccc}
~~Channel   & Index & $M$ & Mixed~~ \\[2ex]
$(D^+_s D^+_s)^1 (3938)$          & 1(S)   & $3978$ & \\
$(D^{*+}_s D^{*+}_s)^1 (4224)$  & 2(S)   & $4232$ & $3978$ \\[2ex]
$(D^+_s D^+_s)^8$          & 3(H)   & $4619$ & \\
$(D^{*+}_s D^{*+}_s)^8$  & 4(H)   & $4636$ & $4377$ \\[2ex]
$(cc)(\bar{s}\bar{s})$      & 5   & $4433$ & \\
$(cc)^*(\bar{s}\bar{s})^*$  & 6   & $4413$ & $4379$ \\[2ex]
$K_1$  & 7   & $4802$ & \\
$K_1$  & 8   & $4369$ & \\
$K_1$  & 9   & $4698$ & \\
$K_1$  & 10   & $4211$ & $4201$ \\[2ex]
$K_2$  & 11   & $4343$ & \\
$K_2$  & 12   & $4753$ & \\
$K_2$  & 13   & $4166$ & \\
$K_2$  & 14   & $4838$ & $4158$ \\[2ex]
$K_3$  & 15   & $4414$ & \\
$K_3$  & 16   & $4427$ & $4373$ \\[2ex]
$K_4$  & 17   & $4413$ & \\
$K_4$  & 18   & $4439$ & $4379$ \\[2ex]
\multicolumn{3}{c}{All of the above channels:} & $3978$
\end{tabular}
\end{ruledtabular}
\end{table}

\begin{table}[!t]
\caption{\label{GresultCC2} The lowest-lying eigen-energies of $cc\bar{s}\bar{s}$ tetraquarks with $IJ^P=01^+$ in the real range calculation. The first column shows the allowed channels and, in the parenthesis, the noninteracting meson-meson threshold value of experiment. Color-singlet (S), hidden-color (H) along with other configurations are indexed in the second column respectively, the third and fourth columns refer to the theoretical mass of each channels and their couplings. (unit: MeV)}
\begin{ruledtabular}
\begin{tabular}{lccc}
~~Channel   & Index & $M$ & Mixed~~ \\[2ex]
$(D^+_s D^{*+}_s)^1 (4081)$      & 1(S)   & $4105$ & \\
$(D^{*+}_s D^{*+}_s)^1 (4224)$  & 2(S)   & $4232$ & $4105$ \\[2ex]
$(D^+_s D^{*+}_s)^8$          & 3(H)   & $4401$ & \\
$(D^{*+}_s D^{*+}_s)^8$     & 4(H)   & $4607$ & $4400$ \\[2ex]
$(cc)^*(\bar{s}\bar{s})^*$   & 5   & $4424$ & $4424$ \\[2ex]
$K_1$  & 6   & $4537$ & \\
$K_1$  & 7   & $4536$ & \\
$K_1$  & 8   & $4528$ & \\
$K_1$  & 9   & $4440$ & \\
$K_1$  & 10   & $4445$ & \\
$K_1$  & 11   & $4371$ & $4305$ \\[2ex]
$K_2$  & 12   & $4417$ & \\
$K_2$  & 13   & $4419$ & \\
$K_2$  & 14   & $4326$ & \\
$K_2$  & 15   & $4699$ & \\
$K_2$  & 16   & $4787$ & \\
$K_2$  & 17   & $4802$ & $4266$ \\[2ex]
$K_3$  & 18   & $4442$ & \\
$K_3$  & 19   & $4443$ & \\
$K_3$  & 20   & $5013$ & $4424$ \\[2ex]
$K_4$  & 21   & $4427$ & \\
$K_4$  & 22   & $4426$ & \\
$K_4$  & 23   & $4953$ & $4423$ \\[2ex]
\multicolumn{3}{c}{All of the above channels:} & $4105$
\end{tabular}
\end{ruledtabular}
\end{table}

\begin{table}[!t]
\caption{\label{GresultCC3} The lowest-lying eigen-energies of $cc\bar{s}\bar{s}$ tetraquarks with $IJ^P=02^+$ in the real range calculation. The first column shows the allowed channels and, in the parenthesis, the noninteracting meson-meson threshold value of experiment. Color-singlet (S), hidden-color (H) along with other configurations are indexed in the second column respectively, the third and fourth columns refer to the theoretical mass of each channels and their couplings. (unit: MeV)}
\begin{ruledtabular}
\begin{tabular}{lccc}
~~Channel   & Index & $M$ & Mixed~~ \\[2ex]
$(D^{*+}_s D^{*+}_s)^1 (4224)$  & 1(S)   & $4232$ & $4232$ \\[2ex]
$(D^{*+}_s D^{*+}_s)^8$  & 2(H)   & $4432$ & $4432$ \\[2ex]
$(cc)^*(\bar{s}\bar{s})^*$  & 3   & $4446$ & $4446$ \\[2ex]
$K_1$  & 4   & $4522$ & \\
$K_1$  & 5   & $4385$ & $4381$ \\[2ex]
$K_2$  & 6   & $4355$ & \\
$K_2$  & 7   & $4666$ & $4354$ \\[2ex]
$K_3$  & 8   & $4448$ & $4448$ \\[2ex]
$K_4$  & 9   & $4446$ & $4446$ \\[2ex]
\multicolumn{3}{c}{All of the above channels:} & $4232$
\end{tabular}
\end{ruledtabular}
\end{table}

\subsection{The $cc\bar{s}\bar{s}$ tetraquarks}

We find only resonances in this sector with quantum numbers $I(J^P)=0(0^+)$ and $0(2^+)$. This result is opposite to the one found in our previous study of $cc\bar{q}\bar{q}$ tetraquarks~\cite{gy:2020dht} and it is related with the ratio between light and heavy quarks that compose the tetraquark system. We shall proceed to discuss below the $J=0$, $1$ and $2$ channels individually.

{\bf The $\bm{I(J^P)=0(0^+)}$ state:}
Two possible meson-meson channels, $D^+_s D^+_s$ and $D^{*+}_s D^{*+}_s$, two diquark-antidiquark channels, $(cc)(\bar{s}\bar{s})$ and $(cc)^*(\bar{s}\bar{s})^*$, along with K-type configurations, are studied first in real-range calculation and our results are shown in Table~\ref{GresultCC1}. The lowest energy level, $(D^+_s D^+_s)^1$, is unbounded and its theoretical mass just equals to the threshold value of two non-interacting $D^+_s$ mesons. This fact is also found in the $(D^{*+}_s D^{*+}_s)^1$ channel whose theoretical mass is 4232 MeV. As for the other exotic configurations, the obtained masses are all higher than the two di-meson channels. In particular, masses of the hidden-color channels are about 4.6 GeV, diquark-antidiquark channels are lower $\sim$4.4 GeV, and the other four K-type configurations are located in the mass interval of 4.2 to 4.8 GeV. Note, too, there is a degeneration between $(cc)^{(*)}(\bar{s}\bar{s})^{(*)}$, $K_3$ and $K_4$ channels around 4.4 GeV.

In a further step, we have performed a coupled-channels calculation on certain configurations, and still no bound states are found. The coupling is quite weak for the color-singlet channels $D^+_s D^+_s$ and $D^{*+}_s D^{*+}_s$. Hidden-color, diquark-antidiquark and K-type structures do not shed any different with respect the color-singlet channel, coupled energies range from 4.2 to 4.4 GeV. These results confirm the nature of scattering states for $D^+_s D^+_s$ and $D^{*+}_s D^{*+}_s$. Moreover, in a complete coupled-channels study, the lowest energy of 3978 MeV for $D^+_s D^+_s$ state is remained. The real-scaling results are consistent with those of $cc\bar{q}\bar{q}$ tetraquarks; however, it is invalid for resonances.

\begin{figure}[ht]
\epsfxsize=3.85in \epsfbox{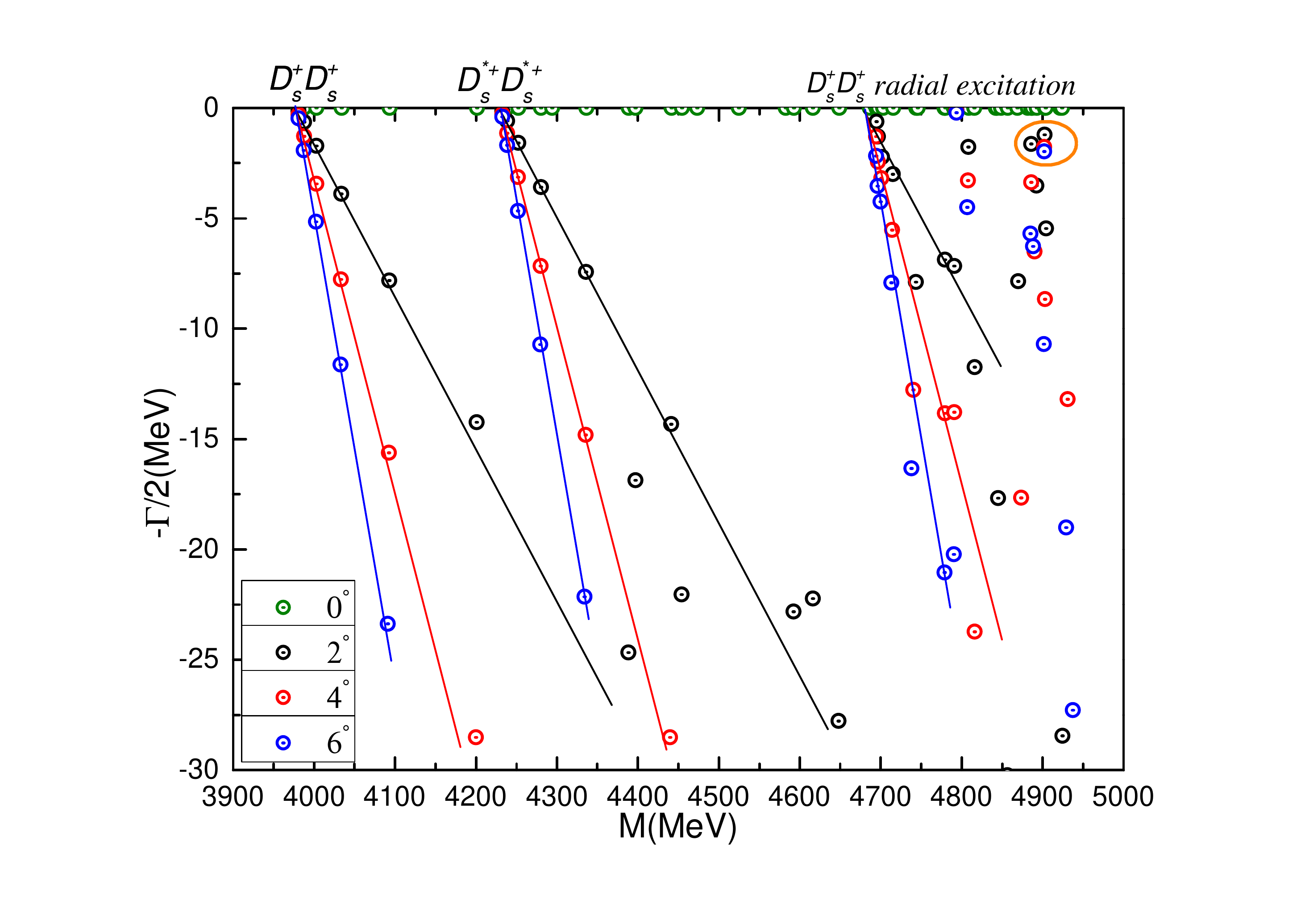}
\vspace*{-1.10cm}
\caption{Complex energies of $cc\bar{s}\bar{s}$ tetraquarks with $IJ^P=00^+$ in the complete coupled channels calculation, $\theta$ varying from $0^\circ$ to $6^\circ$ .} \label{PP1}
\end{figure}

Figure~\ref{PP1} shows the distributions of $cc\bar{s}\bar{s}$ tetraquarks' complex energies in the complete coupled-channels calculation. In the energy gap from 3.9 GeV to 5.0 GeV, most of poles are aligned along the threshold lines. Namely, with the rotated angle $\theta$ varied from $0^\circ$ to $6^\circ$, the $D^{(*)+}_s D^{(*)+}_s$ energy poles always move along with the same color cut lines. However, in the high energy region which is about 0.2 GeV above the $(1S)D^+_s (2S)D^+_s$ threshold, one resonance pole is found. In the yellow circle of Fig.~\ref{PP1}, the three calculated dots with black, red and blue are almost overlapped. This resonance pole has mass and width 4902 MeV and 3.54 MeV, respectively, and it could be identified as a resonance of the $D^+_s D^+_s$ molecular system.

{\bf The $\bm{I(J^P)=0(1^+)}$ state:}
There are two meson-meson channels, $D^+_s D^{*+}_s$ and $D^{*+}_s D^{*+}_s$, one diquark-antidiquark channel, $(cc)^*(\bar{s}\bar{s})^*$, but more K-type channels (18 channels) are allowed due to a much richer combination of color, spin and flavor wave functions which fulfills the Pauli Principle. Table~\ref{GresultCC2} lists the calculated masses of these channels and also their couplings.

Firstly, the situation is similar to the $IJ^P=00^+$ case, \emph{i.e} no bound state is found in the real-range calculation. Secondly, the couplings are extremely weak for both color-singlet and hidden-color channels. In contrast, one finds binding energies for the K-type structures which go from several to hundreds of MeV. The coupled-channels results of these K-type configurations, along with hidden-color and diquark-antidiquark ones, are around 4.4 GeV. Then, after mixing all of the channels listed in Table~\ref{GresultCC2}, the nature of the lowest energy level $D^+_s D^{*+}_s$ is still unchanged, it is a scattering one. Additionally, comparing the results in Table V for $cc\bar{q}\bar{q}$ tetraquarks of our previous work~\cite{gy:2020dht}, one notices that the deeply bound state with $\sim$200 MeV binding energy for $D^+D^{*0}$ is invalid in the $D^+_s D^{*+}_s$ sector.

\begin{figure}[ht]
\epsfxsize=3.85in \epsfbox{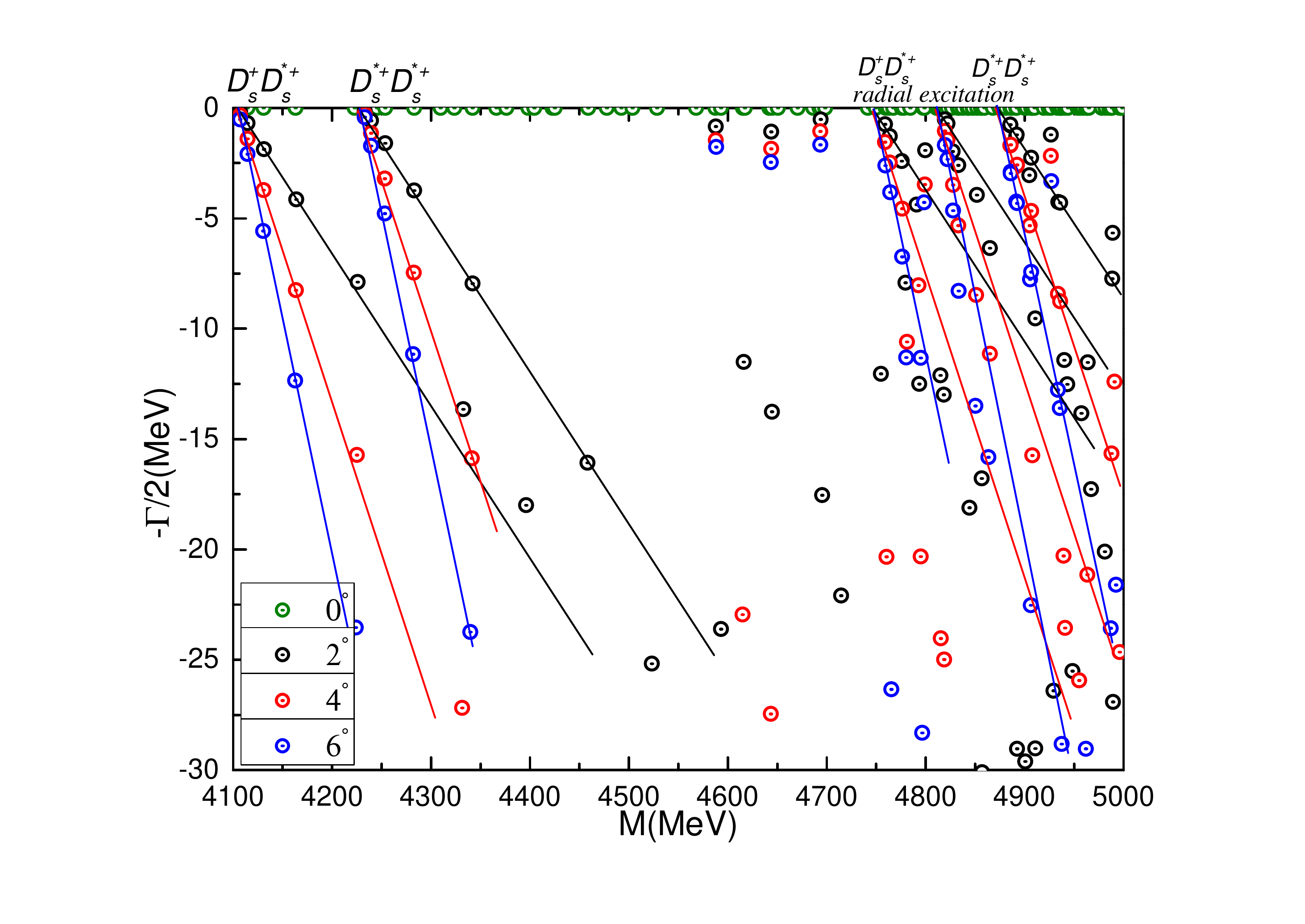}
\vspace*{-1.10cm}
\caption{Complex energies of $cc\bar{s}\bar{s}$ tetraquarks with $IJ^P=01^+$ in the coupled channels calculation, $\theta$ varying from $0^\circ$ to $6^\circ$ .} \label{PP2}
\end{figure}

Our results using the complex scaling method applied to the fully coupled-channels calculation are shown in Fig.~\ref{PP2}. The complex energies of $(1S)D^+_s (1S)D^{*+}_s$, $(1S)D^{*+}_s (1S)D^{*+}_s$ along with their first radial excitation states $(1S)D^+_s (2S)D^{*+}_s$, $(2S)D^+_s (1S)D^{*+}_s$ and $(1S)D^{*+}_s (2S)D^{*+}_s$ are generally aligned along the cut lines when the rotated angle $\theta$ goes from $0^\circ$ to $6^\circ$. Although there are three regions which change slightly in the mass gap 4.55 to 4.70 GeV, the calculated poles still come down gradually when the value of complex angle increases. Hence, neither bound states nor resonances are found within the $IJ^P=01^+$ channel of $cc\bar{s}\bar{s}$ tetraquarks.

{\bf The $\bm{I(J^P)=0(2^+)}$ state:} 
Only one $D^{*+}_s D^{*+}_s$ meson-meson configuration, one $(cc)^*(\bar{s}\bar{s})^*$ diquark-antidiquark structure and six K-type channels contribute to the highest spin channel (see Table~\ref{GresultCC3}). In analogy with the two previous cases, no bound state is obtained neither in each single channel calculation nor in the coupled-channels cases. The mixed results of $K_1$ and $K_2$ types are both around 4.35 GeV, which is lower than those of the other exotic configurations ($\sim4.45\,\text{GeV}$); however, these do not help in realizing a bound state of $D^{*+}_s D^{*+}_s$.

\begin{figure}[ht]
\epsfxsize=3.85in \epsfbox{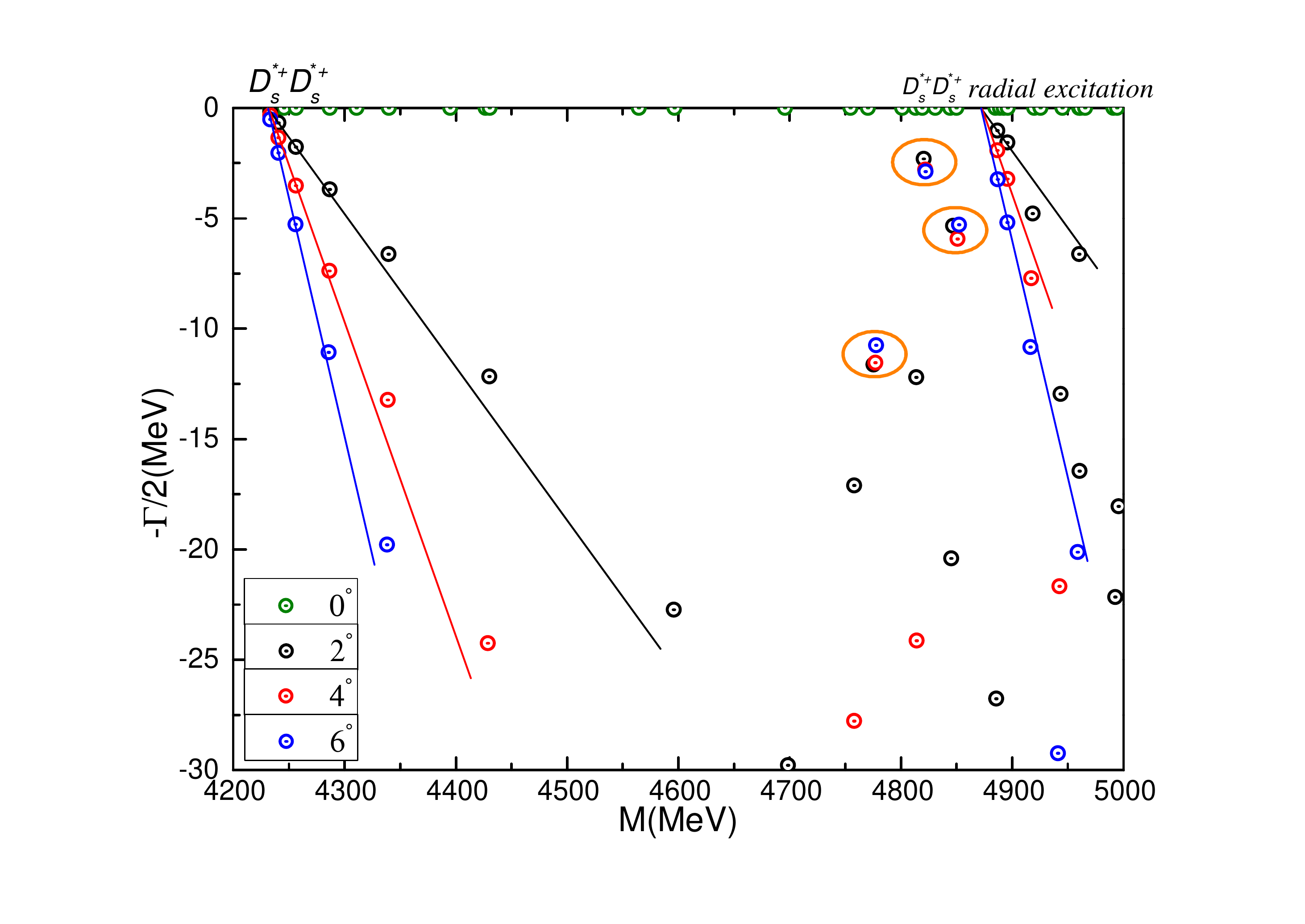}
\vspace*{-1.10cm}
\caption{Complex energies of $cc\bar{s}\bar{s}$ tetraquarks with $IJ^P=02^+$ in the coupled channels calculation, $\theta$ varying from $0^\circ$ to $6^\circ$ .} \label{PP3}
\end{figure}

Nevertheless, thrilling results are found in the complete coupled-channels study by CSM. Figure~\ref{PP3} shows that there are three almost fixed resonance poles between the $(1S)D^{*+}_s (1S)D^{*+}_s$ and $(1S)D^{*+}_s (2S)D^{*+}_s$ threshold lines. Two of them are wide resonances whereas the remaining one is narrow. The $D^{*+}_s D^{*+}_s$ resonances have mass and width $(4821\,\text{MeV},\,5.58\,\text{MeV})$, $(4846\,\text{MeV},\,10.68\,\text{MeV})$ and $(4775\,\text{MeV},\,23.26\,\text{MeV})$, respectively. In the excited energy region which is located about 0.5 GeV higher than the $D^{*+}_s D^{*+}_s$ threshold but 0.1 GeV below its first radial excitation, it is reasonable to find resonances whose nature is of hadronic molecules, and this conclusion has been confirmed by us in study the other multiquark systems~\cite{gy:2020dht, Yang:2015bmv, gy:2020dcp}.

\begin{table}[!t]
\caption{\label{GresultCC4} The lowest-lying eigen-energies of $bb\bar{s}\bar{s}$ tetraquarks with $IJ^P=00^+$ in the real range calculation. The first column shows the allowed channels and, in the parenthesis, the noninteracting meson-meson threshold value of experiment. Color-singlet (S), hidden-color (H) along with other configurations are indexed in the second column respectively, the third and fourth columns refer to the theoretical mass of each channels and their couplings. (unit: MeV)}
\begin{ruledtabular}
\begin{tabular}{lccc}
~~Channel   & Index & $M$ & Mixed~~ \\[2ex]
$(\bar{B}^0_s \bar{B}^0_s)^1 (10734)$          & 1(S)   & $10710$ & \\
$(\bar{B}^*_s \bar{B}^*_s)^1 (10830)$  & 2(S)   & $10800$ & $10710$ \\[2ex]
$(\bar{B}^0_s \bar{B}^0_s)^8$          & 3(H)   & $11184$ & \\
$(\bar{B}^*_s \bar{B}^*_s)^8$  & 4(H)   & $11205$ & $10943$ \\[2ex]
$(bb)(\bar{s}\bar{s})$      & 5   & $10967$ & \\
$(bb)^*(\bar{s}\bar{s})^*$  & 6   & $10901$ & $10896$ \\[2ex]
$K_1$  & 7   & $11445$ & \\
$K_1$  & 8   & $10928$ & \\
$K_1$  & 9   & $11259$ & \\
$K_1$  & 10   & $10863$ & $10843$ \\[2ex]
$K_2$  & 11   & $10877$ & \\
$K_2$  & 12   & $11445$ & \\
$K_2$  & 13   & $10815$ & \\
$K_2$  & 14   & $11441$ & $10802$ \\[2ex]
$K_3$  & 15   & $10902$ & \\
$K_3$  & 16   & $10960$ & $10895$ \\[2ex]
$K_4$  & 17   & $10901$ & \\
$K_4$  & 18   & $10980$ & $10897$ \\[2ex]
\multicolumn{3}{c}{All of the above channels:} & $10710$
\end{tabular}
\end{ruledtabular}
\end{table}

\begin{table}[!t]
\caption{\label{GresultCC5} The lowest-lying eigen-energies of $bb\bar{s}\bar{s}$ tetraquarks with $IJ^P=01^+$ in the real range calculation. The first column shows the allowed channels and, in the parenthesis, the noninteracting meson-meson threshold value of experiment. Color-singlet (S), hidden-color (H) along with other configurations are indexed in the second column respectively, the third and fourth columns refer to the theoretical mass of each channels and their couplings. (unit: MeV)}
\begin{ruledtabular}
\begin{tabular}{lccc}
~~Channel   & Index & $M$ & Mixed~~ \\[2ex]
$(\bar{B}^0_s \bar{B}^*_s)^1 (10782)$      & 1(S)   & $10755$ & \\
$(\bar{B}^*_s \bar{B}^*_s)^1 (10830)$  & 2(S)   & $10800$ & $10755$ \\[2ex]
$(\bar{B}^0_s \bar{B}^*_s)^8$          & 3(H)   & $10949$ & \\
$(\bar{B}^*_s \bar{B}^*_s)^8$     & 4(H)   & $11185$ & $10949$ \\[2ex]
$(bb)^*(\bar{s}\bar{s})^*$   & 5   & $10906$ & $10906$ \\[2ex]
$K_1$  & 6   & $11041$ & \\
$K_1$  & 7   & $11048$ & \\
$K_1$  & 8   & $11038$ & \\
$K_1$  & 9   & $10936$ & \\
$K_1$  & 10   & $10949$ & \\
$K_1$  & 11   & $10917$ & $10870$ \\[2ex]
$K_2$  & 12   & $10911$ & \\
$K_2$  & 13   & $10914$ & \\
$K_2$  & 14   & $10879$ & \\
$K_2$  & 15   & $11216$ & \\
$K_2$  & 16   & $11483$ & \\
$K_2$  & 17   & $11373$ & $10840$ \\[2ex]
$K_3$  & 18   & $10928$ & \\
$K_3$  & 19   & $10929$ & \\
$K_3$  & 20   & $11557$ & $10907$ \\[2ex]
$K_4$  & 21   & $10911$ & \\
$K_4$  & 22   & $10908$ & \\
$K_4$  & 23   & $11458$ & $10906$ \\[2ex]
\multicolumn{3}{c}{All of the above channels:} & $10755$
\end{tabular}
\end{ruledtabular}
\end{table}

\begin{table}[!t]
\caption{\label{GresultCC6} The lowest-lying eigen-energies of $bb\bar{s}\bar{s}$ tetraquarks with $IJ^P=02^+$ in the real range calculation. The first column shows the allowed channels and, in the parenthesis, the noninteracting meson-meson threshold value of experiment. Color-singlet (S), hidden-color (H) along with other configurations are indexed in the second column respectively, the third and fourth columns refer to the theoretical mass of each channels and their couplings. (unit: MeV)}
\begin{ruledtabular}
\begin{tabular}{lccc}
~~Channel   & Index & $M$ & Mixed~~ \\[2ex]
$(\bar{B}^*_s \bar{B}^*_s)^1 (10830)$  & 1(S)   & $10800$ & $10800$ \\[2ex]
$(\bar{B}^*_s \bar{B}^*_s)^8$  & 2(H)   & $10959$ & $10959$ \\[2ex]
$(bb)^*(\bar{s}\bar{s})^*$  & 3   & $10915$ & $10915$ \\[2ex]
$K_1$  & 4   & $11023$ & \\
$K_1$  & 5   & $10894$ & $10879$ \\[2ex]
$K_2$  & 6   & $10870$ & \\
$K_2$  & 7   & $11186$ & $10869$ \\[2ex]
$K_3$  & 8   & $10918$ & $10918$ \\[2ex]
$K_4$  & 9   & $10916$ & $10916$ \\[2ex]
\multicolumn{3}{c}{All of the above channels:} & $10800$
\end{tabular}
\end{ruledtabular}
\end{table}

\subsection{The $bb\bar{s}\bar{s}$ tetraquarks}

We proceed here to analyze the $bb\bar{s}\bar{s}$ tetraquark system with quantum numbers $I(J^P)=0(0^+)$, $0(1^+)$ and $0(2^+)$. The situation is similar to the $cc\bar{s}\bar{s}$ case, with only narrow resonances found in the $I(J^P)=0(0^+)$ and $0(2^+)$ channels; meanwhile, this result is also in contrast with the one obtained for $bb\bar{q}\bar{q}$ tetraquarks~\cite{gy:2020dht}. The details are as following.

{\bf The $\bm{I(J^P)=0(0^+)}$ state:}
Table~\ref{GresultCC4} summarizes all possible channels for the $I(J^P)=0(0^+)$ $bb\bar{s}\bar{s}$ tetraquark. In particular, there are two meson-meson channels $\bar{B}^0_s \bar{B}^0_s$ and $\bar{B}^*_s \bar{B}^*_s$, both color-singlet and hidden-color channels are considered. Moreover, there are two diquark-antidiquark structures, $(bb)(\bar{s}\bar{s})$ and $(bb)^*(\bar{s}\bar{s})^*$, and 12 K-type channels. The calculated mass of each single channel ranges 10.7 to 11.5 GeV, and no bound state is observed. Additionally, after coupling between the same kind of configurations, one can conclude that the coupling is weak in di-meson case and it is quite comparable among diquark-antidiquark and K-type structures.

The nature of scattering for lowest state $\bar{B}^0_s \bar{B}^0_s$ remains in the complete coupled-channels calculation with rotated angle $\theta=0^\circ$. However, three narrow resonance states are obtained in the complex-scaling analysis. In the mass region from 10.7 to 11.5 GeV, Fig.~\ref{PP4} established the complex energy distributions of $\bar{B}^0_s \bar{B}^0_s$, $\bar{B}^*_s \bar{B}^*_s$ and their first radial excitation states. There are two orange circles which surround the resonance poles whose masses and widths are (11.31 GeV, 1.86 MeV), (11.33 GeV, 1.84 MeV) and (11.41 GeV, 1.54 MeV), respectively. The first two resonances can be identified as $\bar{B}^*_s \bar{B}^*_s$ molecular states because they are $\sim0.5\,GeV$ higher than its threshold value and the third one should be interpreted as a $\bar{B}^0_s \bar{B}^0_s$ because its location is just between $(1S)\bar{B}^0_s (2S)\bar{B}^0_s$ and $(1S)\bar{B}^*_s (2S)\bar{B}^*_s$ states. Finally, after comparing our results of $cc\bar{s}\bar{s}$ and $bb\bar{s}\bar{s}$ tetraquarks, we conclude that, with much heavier constituent quark components included, more narrow molecular resonances will be found around 0.2 GeV interval near the first radial excitation states.

{\bf The $\bm{I(J^P)=0(1^+)}$ state:}
The results listed in Table~\ref{GresultCC5} highlight that tightly bound and narrow resonance states obtained in $bb\bar{q}\bar{q}$ tetraquarks~\cite{gy:2020dht} are not found in this case. Firstly, the lowest channel $\bar{B}^0_s \bar{B}^*_s$ is of scattering nature both in single channel calculation and coupled-channels one. Secondly, the mass of diquark-antidiquark configuration is higher than meson-meson channels and its value of 10.91 GeV is very close to the hidden-color channels, 10.95 GeV. Furthermore, the other K-type configurations produce masses slightly lower ($\sim10.90\,GeV$) than the former case.

Figure~\ref{PP5} shows that the scattering nature of $\bar{B}^0_s \bar{B}^*_s$ and $\bar{B}^*_s \bar{B}^*_s$ states is even clearer when the CSM is employed. More specifically, in the mass interval from 10.7 to 11.3 GeV, the calculated complex energies always move along with the varied angle $\theta$. There is no fixed pole in the energy region which is around 0.6 GeV above the $\bar{B}^0_s \bar{B}^*_s$ threshold. This fact is consistent with the $cc\bar{s}\bar{s}$ results discussed above.

{\bf The $\bm{I(J^P)=0(2^+)}$ state:}
For the highest spin channel of $bb\bar{s}\bar{s}$ tetraquarks, Table~\ref{GresultCC6} summarizes our theoretical findings in real-range method. Among our results, the following are of particular interest: (i) only one di-meson channel $\bar{B}^*_s \bar{B}^*_s$ exists and it is unbounded if we consider either the single channel or multi-channels coupling calculation, and (ii) the other exotic configurations which include hidden-color, diquark-antidiquark and K-type are all excited states with masses on 10.9 GeV.

In a further step, in which the complex analysis is adopted, three resonances are obtained. It is quite obvious in Fig.~\ref{PP6} that three fixed poles, marked with orange circles, are located at around 11.35 GeV and near the real-axis. The exact masses and widths of these $\bar{B}^*_s \bar{B}^*_s$ resonances are (11.33 GeV, 1.48 MeV), (11.36 GeV, 4.18 MeV) and (11.41 GeV, 2.52 MeV), respectively. These found narrow resonances are 0.6 GeV above $\bar{B}^*_s \bar{B}^*_s$ threshold, meanwhile they approach its first radial excitation state.

%%%%%%%%%%
\begin{table}[!t]
\caption{\label{GresultCC7} The lowest-lying eigen-energies of $cb\bar{s}\bar{s}$ tetraquarks with $IJ^P=00^+$ in the real range calculation. The first column shows the allowed channels and, in the parenthesis, the noninteracting meson-meson threshold value of experiment. Color-singlet (S), hidden-color (H) along with other configurations are indexed in the second column respectively, the third and fourth columns refer to the theoretical mass of each channels and their couplings. (unit: MeV)}
\begin{ruledtabular}
\begin{tabular}{lccc}
~~Channel   & Index & $M$ & Mixed~~ \\[2ex]
$(D^+_s \bar{B}^0_s)^1 (7336)$          & 1(S)   & $7344$ & \\
$(D^{*+}_s \bar{B}^*_s)^1 (7527)$  & 2(S)   & $7516$ & $7344$ \\[2ex]
$(D^+_s \bar{B}^0_s)^8$          & 3(H)   & $7910$ & \\
$(D^{*+}_s \bar{B}^*_s)^8$  & 4(H)   & $7927$ & $7678$ \\[2ex]
$(cb)(\bar{s}\bar{s})$      & 5   & $7726$ & \\
$(cb)^*(\bar{s}\bar{s})^*$  & 6   & $7675$ & $7662$ \\[2ex]
$K_1$  & 7   & $8171$ & \\
$K_1$  & 8   & $8274$ & \\
$K_1$  & 9   & $8369$ & \\
$K_1$  & 10   & $8145$ & $7613$ \\[2ex]
$K_2$  & 11   & $7896$ & \\
$K_2$  & 12   & $8266$ & \\
$K_2$  & 13   & $7758$ & \\
$K_2$  & 14   & $8282$ & $7629$ \\[2ex]
$K_3$  & 15   & $8647$ & \\
$K_3$  & 16   & $8181$ & \\
$K_3$  & 17   & $8321$ & \\
$K_3$  & 18   & $8675$ & $8010$ \\[2ex]
$K_4$  & 19   & $8199$ & \\
$K_4$  & 20   & $8359$ & $8063$ \\[2ex]
\multicolumn{3}{c}{All of the above channels:} & $7344$
\end{tabular}
\end{ruledtabular}
\end{table}

\begin{figure}[ht]
\epsfxsize=3.85in \epsfbox{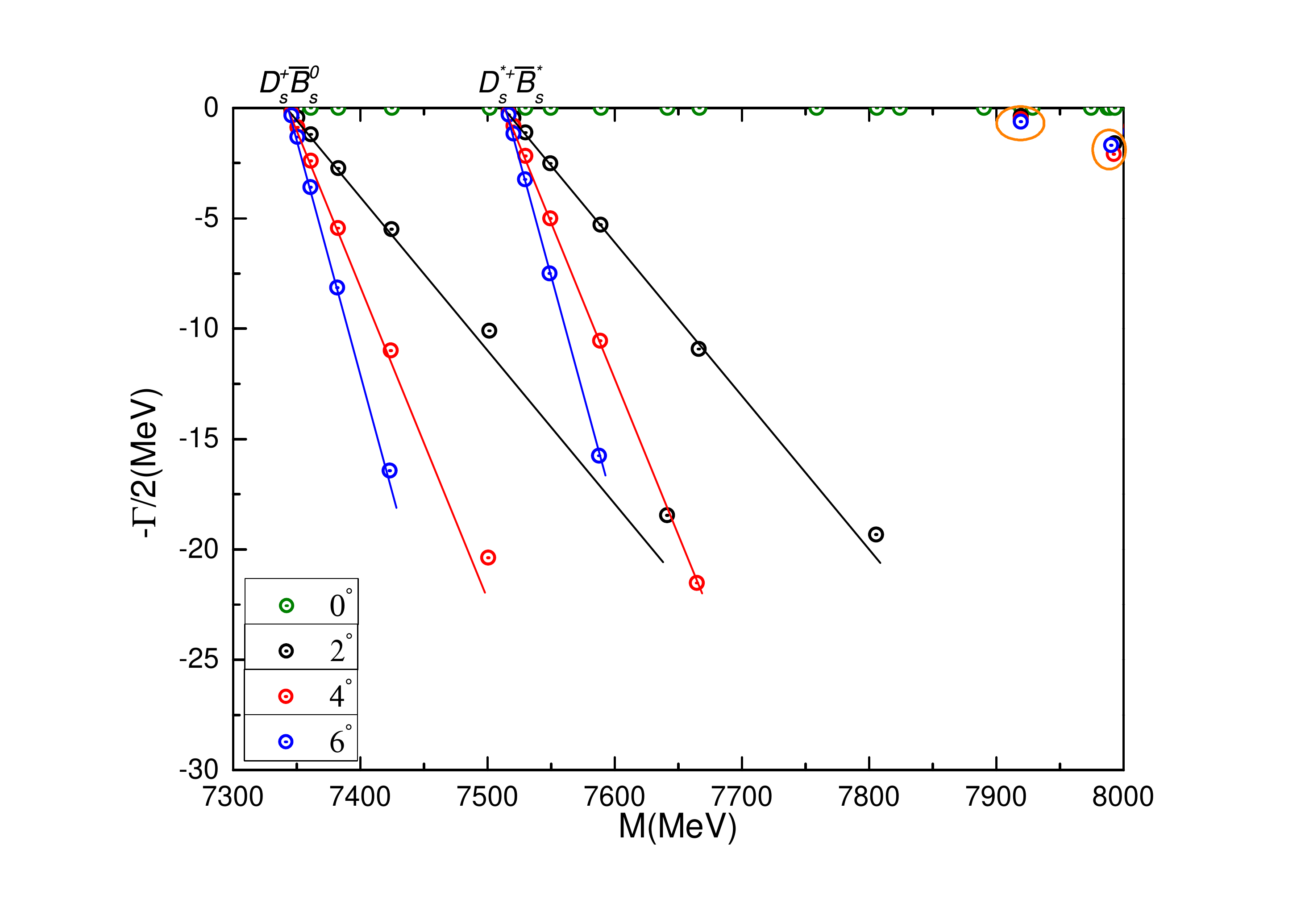}
\vspace*{-1.10cm}
\caption{Complex energies of $cb\bar{s}\bar{s}$ tetraquarks with $IJ^P=00^+$ in the coupled channels calculation, $\theta$ varying from $0^\circ$ to $6^\circ$ .} \label{PP7}
\end{figure}

\subsection{The $cb\bar{s}\bar{s}$ tetraquarks}

Several narrow resonances are found in this sector with quantum numbers $I(J^P)=0(0^+)$, $0(1^+)$ and $0(2^+)$. However, no bound states are found as in the case of $cb\bar{q}\bar{q}$ tetraquarks~\cite{gy:2020dht}.

{\bf The $\bm{I(J^P)=0(0^+)}$ channel:}
There are two meson-meson channels, $D^+_s \bar{B}^0_s$ and $D^{*+}_s \bar{B}^*_s$, two diquark-antidiquark structures, $(cb)(\bar{s}\bar{s})$ and $(cb)^*(\bar{s}\bar{s})^*$, and 14 K-type channels (see Table\ref{GresultCC7}). The single channel calculation produces masses which ranges from 7.34 to 8.67 GeV, and all states are scattering ones. The coupled-channels study for each kind of structure reveals weak couplings in di-meson configuration of color-singlet channels and stronger ones for the other structures, with masses above 7.6 GeV.

If we now rotate the angle $\theta$ from $0^\circ$ to $6^\circ$ in a fully coupled-channels calculation, Fig.~\ref{PP7} shows the distribution of complex energy points of $D^+_s \bar{B}^0_s$ and $D^{*+}_s \bar{B}^*_s$. It is obvious to notice that there are two stable poles in the mass region from 7.3 to 8.0 GeV. Actually, their calculated masses and widths are (7.92 GeV, 1.02 MeV) and (7.99 GeV, 3.22 MeV), respectively. Because they are much more close to the $D^{*+}_s \bar{B}^*_s$ threshold lines, the two narrow resonances can be identified as $D^{*+}_s \bar{B}^*_s$ molecules. 

\begin{table}[!t]
\caption{\label{GresultCC8} The lowest-lying eigen-energies of $cb\bar{s}\bar{s}$ tetraquarks with $IJ^P=01^+$ in the real range calculation. The first column shows the allowed channels and, in the parenthesis, the noninteracting meson-meson threshold value of experiment. Color-singlet (S), hidden-color (H) along with other configurations are indexed in the second column respectively, the third and fourth columns refer to the theoretical mass of each channels and their couplings. (unit: MeV)}
\begin{ruledtabular}
\begin{tabular}{lccc}
~~Channel   & Index & $M$ & Mixed~~ \\[2ex]
$(D^+_s \bar{B}^*_s)^1 (7384)$          & 1(S)   & $7389$ & \\
$(D^{*+}_s \bar{B}^0_s)^1 (7479)$     & 2(S)   & $7471$ & \\
$(D^{*+}_s \bar{B}^*_s)^1 (7527)$  & 3(S)   & $7516$ & $7389$ \\[2ex]
$(D^+_s \bar{B}^*_s)^8$          & 4(H)   & $7900$ & \\
$(D^{*+}_s \bar{B}^0_s)^8$     & 5(H)   & $7891$ & \\
$(D^{*+}_s \bar{B}^*_s)^8$      & 6(H)   & $7920$ & $7684$ \\[2ex]
$(cb)(\bar{s}\bar{s})^*$      & 7   & $7683$ & \\
$(cb)^*(\bar{s}\bar{s})$      & 8   & $7680$ & \\
$(cb)^*(\bar{s}\bar{s})^*$  & 9   & $7725$ & $7671$ \\[2ex]
$K_1$  & 10   & $7796$ & \\
$K_1$  & 11   & $8172$ & \\
$K_1$  & 12   & $8009$ & \\
$K_1$  & 13   & $7695$ & \\
$K_1$  & 14   & $7760$ & \\
$K_1$  & 15   & $7634$ & $7620$ \\[2ex]
$K_2$  & 16   & $7607$ & \\
$K_2$  & 17   & $7621$ & \\
$K_2$  & 18   & $7510$ & \\
$K_2$  & 19   & $8137$ & \\
$K_2$  & 20   & $8211$ & \\
$K_2$  & 21   & $8209$ & $7505$ \\[2ex]
$K_3$  & 22   & $7705$ & \\
$K_3$  & 23   & $7706$ & \\
$K_3$  & 24   & $7682$ & \\
$K_3$  & 25   & $7734$ & \\
$K_3$  & 26   & $7733$ & \\
$K_3$  & 27   & $8298$ & $7666$ \\[2ex]
$K_4$  & 28   & $7687$ & \\
$K_4$  & 29   & $7677$ & \\
$K_4$  & 30   & $7771$ & $7670$ \\[2ex]
\multicolumn{3}{c}{All of the above channels:} & $7389$
\end{tabular}
\end{ruledtabular}
\end{table}

\begin{figure}[ht]
\epsfxsize=3.85in \epsfbox{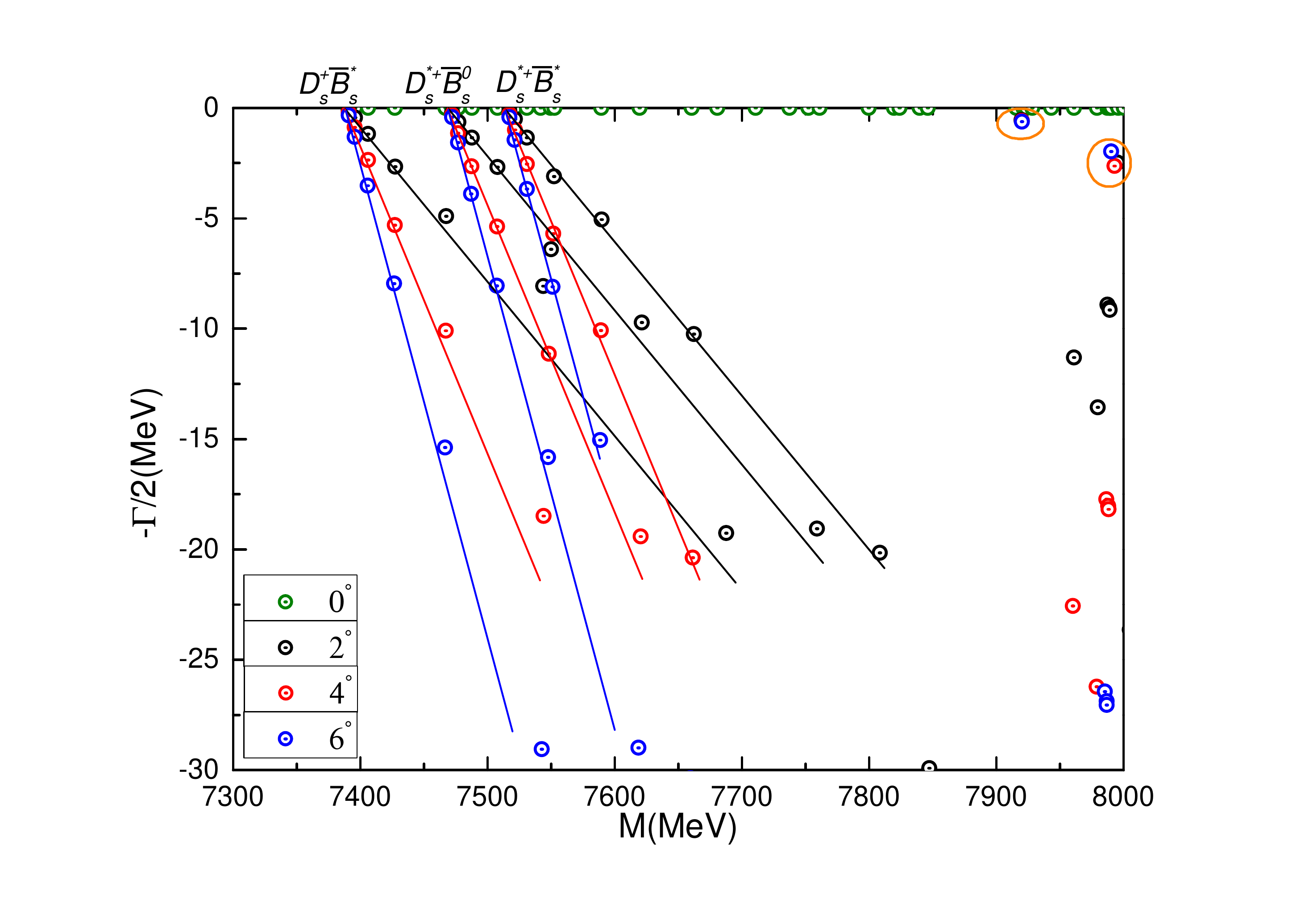}
\vspace*{-1.10cm}
\caption{Complex energies of $cb\bar{s}\bar{s}$ tetraquarks with $IJ^P=01^+$ in the coupled channels calculation, $\theta$ varying from $0^\circ$ to $6^\circ$ .} \label{PP8}
\end{figure}

{\bf The $\bm{I(J^P)=0(1^+)}$ channel:}
There are 30 possible channels in this case and they are listed in Table~\ref{GresultCC8}; in particular, one has three meson-meson channels: $D^+_s \bar{B}^*_s$, $D^{*+}_s \bar{B}^0_s$ and $D^{*+}_s \bar{B}^*_s$, the diquark-antidiquark channels $(cb)(\bar{s}\bar{s})^*$, $(cb)^*(\bar{s}\bar{s})$ and $(cb)^*(\bar{s}\bar{s})^*$, and the remaining 21 channels are of K-type configurations. In our first kind of calculation, the single channel masses are located in the energy interval 7.39 to 8.23 GeV. Particularly, the color-singlet channels of di-meson configurations present masses which are below 7.52 GeV, and the other exotic structures' are above this level. Then, in the coupled-channels computation, the lowest energy of color-singlet channels is still the $D^+_s \bar{B}^*_s$ threshold value, 7.39 GeV. Masses of the other configurations are about 7.67 GeV, except for 7.51 GeV of $K_2$-type channels. %Although bound state is till not obtained in the complete coupled-channels calculation, two narrow resonances are found then in the complex range.

Figure~\ref{PP8} depicts mostly distributions of scattering states of $D^+_s \bar{B}^*_s$, $D^{*+}_s \bar{B}^0_s$ and $D^{*+}_s \bar{B}^*_s$, \emph{i.e} the calculated complex energies are basically aligned along their respective threshold lines. However, two stable poles are located in the top right corner of this figure. Inside the two orange circles, one can find that the black, red and blue dots (which are the results of $2^\circ$, $4^\circ$ and $6^\circ$ rotated angle, respectively) almost overlap. Together with the fact that near $D^{*+}_s \bar{B}^*_s$ threshold lines appear, they can be identified as $D^{*+}_s$ and $\bar{B}^*_s$ resonances whose masses and widths are (7.92 GeV, 1.20 MeV) and (7.99 GeV, 4.96 MeV), respectively. 

\begin{table}[!t]
\caption{\label{GresultCC9} The lowest-lying eigen-energies of $cb\bar{s}\bar{s}$ tetraquarks with $IJ^P=02^+$ in the real range calculation. The first column shows the allowed channels and, in the parenthesis, the noninteracting meson-meson threshold value of experiment. Color-singlet (S), hidden-color (H) along with other configurations are indexed in the second column respectively, the third and fourth columns refer to the theoretical mass of each channels and their couplings. (unit: MeV)}
\begin{ruledtabular}
\begin{tabular}{lccc}
~~Channel   & Index & $M$ & Mixed~~ \\[2ex]
$(D^{*+}_s \bar{B}^*_s)^1 (7527)$  & 1(S)   & $7516$ & $7516$ \\[2ex]
$(D^{*+}_s \bar{B}^*_s)^8$  & 2(H)   & $7712$ & $7712$ \\[2ex]
$(cb)^*(\bar{s}\bar{s})^*$  & 3   & $7698$ & $7698$ \\[2ex]
$K_1$  & 4   & $7804$ & \\
$K_1$  & 5   & $7705$ & $7704$ \\[2ex]
$K_2$  & 6   & $7624$ & \\
$K_2$  & 7   & $8205$ & $7622$ \\[2ex]
$K_3$  & 8   & $8311$ & \\
$K_3$  & 9   & $7701$ & $7696$ \\[2ex]
$K_4$  & 10   & $7697$ & $7697$ \\[2ex]
\multicolumn{3}{c}{All of the above channels:} & $7516$
\end{tabular}
\end{ruledtabular}
\end{table}

\begin{figure}[ht]
\epsfxsize=3.85in \epsfbox{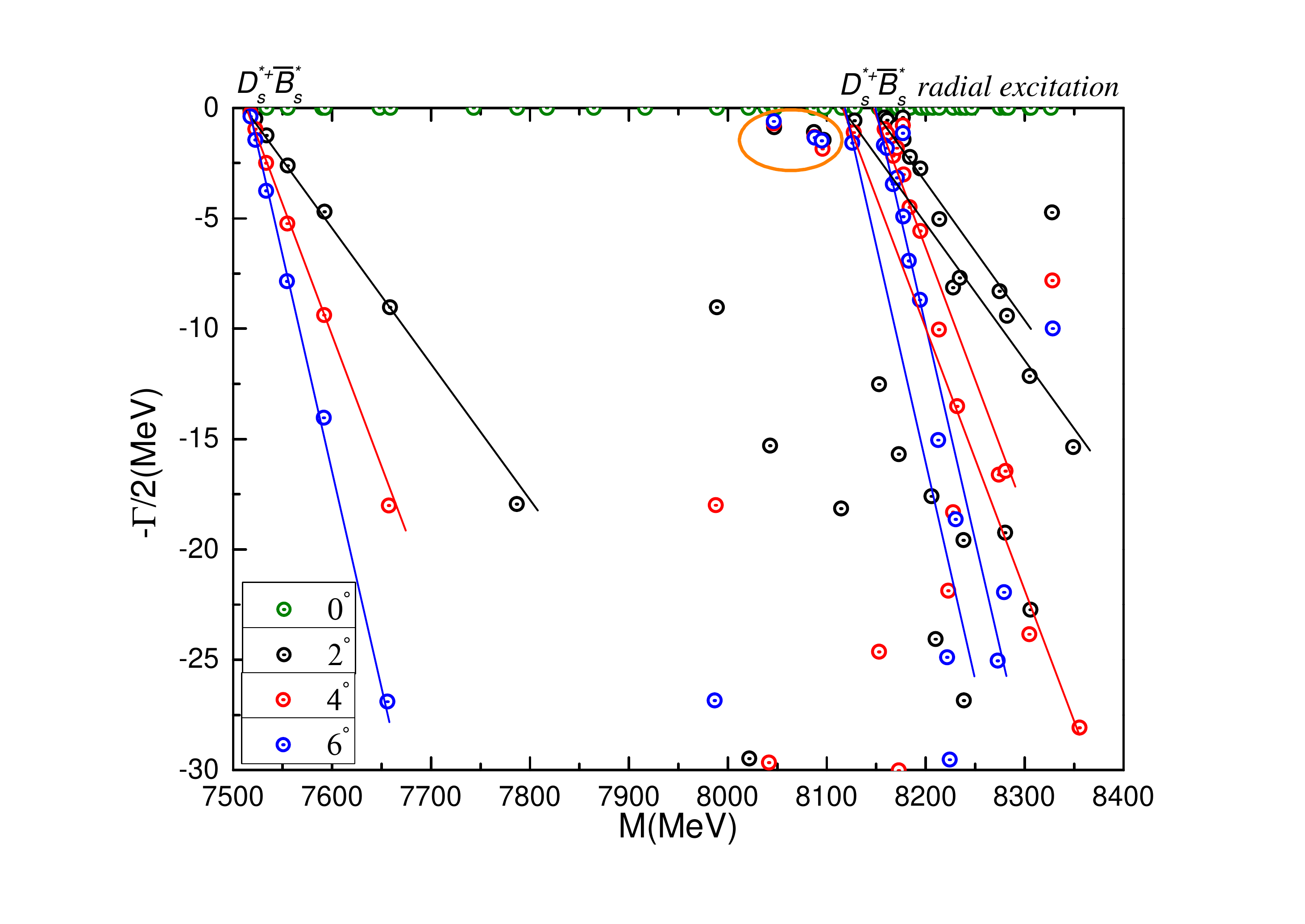}
\vspace*{-1.10cm}
\caption{Complex energies of $cb\bar{s}\bar{s}$ tetraquarks with $IJ^P=02^+$ in the coupled channels calculation, $\theta$ varying from $0^\circ$ to $6^\circ$ .} \label{PP9}
\end{figure}

{\bf The $\bm{I(J^P)=0(2^+)}$ channel:}
Only one meson-meson channel, $D^{*+}_s \bar{B}^*_s$, one diquark-antidiquark channel, $(cb)^*(\bar{s}\bar{s})^*$, and 7 K-type configurations must be considered in this case. Their calculated masses are listed in Table~\ref{GresultCC9}. As all other cases discussed above, no bound states are found neither in the single channel computation nor in the coupled-channels case. The lowest scattering state of $D^{*+}_s \bar{B}^*_s$ is located at 7.52 GeV and all other excited states, in coupled-channels calculation, are below 7.72 GeV.

In contrast to the $cb\bar{q}\bar{q}$ tetraquarks~\cite{gy:2020dht}, two $cb\bar{s}\bar{s}$ resonances are found in the complete coupled-channels calculation when complex range method is used. Figure~\ref{PP9} shows an orange circle, which is near the threshold lines $(1S)D^{*+}_s (2S)\bar{B}^*_s$ and $(2S)D^{*+}_s (1S)\bar{B}^*_s$, surrounding two fixed resonance poles. The calculated masses and widths are (8.05 GeV, 1.42 MeV) and (8.10 GeV, 2.90 MeV), respectively. Apparently, these two narrow $D^{*+}_s \bar{B}^*_s$ resonances are $\sim 0.6\,\text{GeV}$ higher than their threshold and this is just similar to our previous results.

\begin{table}[!t]
\caption{\label{GresultCCT} Possible resonance states of $QQ\bar{s}\bar{s}$ $(Q=c, b)$ tetraquarks. (unit: MeV)}
\begin{ruledtabular}
\begin{tabular}{lccc}
~~ $IJ^P$ & Resonance   & Mass & Width~~ \\[2ex]
~~$00^+$  & $D^+_s D^+_s$   & 4902 & 3.54\\
 & $\bar{B}^*_s \bar{B}^*_s$   & 11306 & 1.86\\
 & $\bar{B}^*_s \bar{B}^*_s$   & 11333 & 1.84\\
 & $\bar{B}^0_s \bar{B}^0_s$   & 11412 & 1.54\\
 & $D^{*+}_s \bar{B}^*_s$   & 7919 & 1.02\\
 & $D^{*+}_s \bar{B}^*_s$   & 7993 & 3.22\\[2ex]
~~$01^+$  & $D^{*+}_s \bar{B}^*_s$   & 7920 & 1.20\\
 & $D^{*+}_s \bar{B}^*_s$   & 7995 & 4.96\\[2ex]
~~$02^+$  & $D^{*+}_s D^{*+}_s$   & 4821 & 5.58\\
 & $D^{*+}_s D^{*+}_s$   & 4846 & 10.68\\
 & $D^{*+}_s D^{*+}_s$   & 4775 & 23.26\\
 & $\bar{B}^*_s \bar{B}^*_s$   & 11329 & 1.48\\
 & $\bar{B}^*_s \bar{B}^*_s$   & 11356 & 4.18\\
 & $\bar{B}^*_s \bar{B}^*_s$   & 11410 & 2.52\\
 & $D^{*+}_s \bar{B}^*_s$   & 8046 & 1.42\\
 & $D^{*+}_s \bar{B}^*_s$   & 8096 & 2.90\\
\end{tabular}
\end{ruledtabular}
\end{table}

%%%%%%%%%%%%%%%%%%%%%%%%%%%%%%%%%%%%%%%%%%%%%%%%%%%%%%%%%%%%%%%%%%%%%%%%%%%%%%%%

\section{Epilogue}
\label{sec:summary}

The $QQ\bar{s}\bar{s}$ tetraquarks with spin-parity $J^P=0^+$, $1^+$ and $2^+$, and in the isoscalar sector $I=0$ have been systemically investigated. This is a natural extension of our previous work on double-heavy tetraquarks $QQ\bar{q}\bar{q}$ $(q=u, d)$; however, not only the meson-meson and diquark-antidiquark configurations, with their allowed color structures: color-singlet, hidden-color, color triplet-antitriplet and color sextet-antisextet, are considered but also four K-type configurations are included herein.

The chiral quark model contains the perturbative one-gluon exchange interaction and the nonperturbative linear-screened confinement and Goldstone-boson exchange interactions between anti-strange quarks. This model has been successfully applied to the description of hadron, hadron-hadron and multiquark phenomenology. In order to distinguish among bound states, resonances and scattering poles the complex scaling method is used. Following Ref.~\cite{Hiyama:2003cu}, we employ Gaussian trial functions with ranges  in geometric progression. This enables the optimization of ranges employing a small number of free parameters.

For the three types of tetraquarks: $cc\bar{s}\bar{s}$, $bb\bar{s}\bar{s}$ and $cb\bar{s}\bar{s}$, no bound state is found in any quantum-number channel studied herein, and this is in contrast with the $QQ\bar{q}\bar{q}$ sector. However, several resonances are available with different quantum numbers and nature. Table~\ref{GresultCCT} collects our results showing the mass and width of each found resonance. Some details of such resonances are summarized below.

All found resonances are about 0.6 GeV higher than their corresponding threshold and near the first radial excitation states, around 0.2 GeV energy region. For the $cc\bar{s}\bar{s}$ tetraquark, one narrow $D^+_s D^+_s$ resonance is obtained in $IJ^P=00^+$ channel with mass and width 4.9 GeV and 3.54 MeV, respectively. Besides, another narrow resonance of 5.58 MeV width and two wide ones with widths of 10.68 MeV and 23.26 MeV are found for $D^{*+}_s D^{*+}_s$ in the $IJ^P=02^+$ channel; their masses are 4.82 GeV, 4.85 GeV and 4.78 GeV, respectively.
 
Similarly to $cc\bar{s}\bar{s}$ tetraquarks, narrow resonances are only found in $00^+$ and $02^+$ states for $bb\bar{s}\bar{s}$ sector. However, with much heavier flavor quarks included, more resonances are available. Specifically, there are two $\bar{B}^*_s\bar{B}^*_s$ and one $\bar{B}^0_s\bar{B}^0_s$ resonances in $IJ^P=00^+$ channel. Their masses and widths are (11.31 GeV, 1.86 MeV), (11.33 GeV, 1.84 GeV) and (11.41 GeV, 1.54 MeV), respectively. Meanwhile, in the $IJ^P=02^+$ channel, three $\bar{B}^*_s\bar{B}^*_s$ resonances are obtained with masses and widths (11.33 GeV, 1.48 MeV), (11.36 GeV, 4.18 GeV) and (11.41 GeV, 2.52 MeV), respectively.

Furthermore, two $D^{*+}_s\bar{B}^*_s$ narrow resonances have been found in each $IJ^P=00^+$, $01^+$ and $02^+$ channel. Their masses and widths can be summarized as follows: $D^{*+}_s\bar{B}^*_s$(7.92 GeV, 1.02 MeV) and $D^{*+}_s\bar{B}^*_s$(7.99 GeV, 3.22 MeV) within the $IJ^P=00^+$ channel; $D^{*+}_s\bar{B}^*_s$(7.92 GeV, 1.20 MeV) and $D^{*+}_s\bar{B}^*_s$(7.99 GeV, 4.96 MeV) within the $IJ^P=01^+$ channel; and $D^{*+}_s\bar{B}^*_s$(8.05 GeV, 1.42 MeV) and $D^{*+}_s\bar{B}^*_s$(8.09 GeV, 2.90 MeV) in the case of $IJ^P=02^+$.

%%%%%%%%%%%%%%%%%%%%%%%%%%%%%%%%%%%%%%%%%%%%%%%%%%%%%%%%%%%%%%%%%%%%%%%%%%%%%%%%

\begin{acknowledgments}
Work partially financed by: the National Natural Science Foundation of China under Grant No. 11535005 and No. 11775118; the Ministerio Espa\~nol de Ciencia e Innovaci\'on under grant No. PID2019-107844GB-C22; and the Junta de Andaluc\'ia under contract No. Operativo FEDER Andaluc\'ia 2014-2020 UHU-1264517.
\end{acknowledgments}

%%%%%%%%%%%%%%%%%%%%%%%%%%%%%%%%%%%%%%%%%%%%%%%%%%%%%%%%%%%%%%%%%%%%%%%%%%%%%%%%

\bibliography{QQsstetraquarks}

\end{document}